\documentclass{aa}
\usepackage{epsfig}
\usepackage{graphicx}
\usepackage{amssymb}
\usepackage[below]{placeins}
\usepackage{natbib}
\bibpunct{(}{)}{;}{a}{}{,}

\def \arcmin{$^{\prime}$}
\def \sncc{SN$_{\mathrm{CC}}$}

\begin{document}

\title{Chemical enrichment in the cluster of galaxies Hydra~A}
\author{A. Simionescu\inst{1}
 \and N. Werner\inst{2,6} 
 \and H. B\"ohringer\inst{1}
 \and J. S. Kaastra \inst{2}
 \and A. Finoguenov \inst{1,3} 
 \and M. Br\"uggen \inst{4}
 \and P. E. J. Nulsen \inst{5} }

\institute{     Max-Planck-Institute for Extraterrestrial Physics, Giessenbachstr., 85748, Garching, Germany		
\and		SRON Netherlands Institute for Space Research, Sorbonnelaan 2, NL - 3584 CA Utrecht, the Netherlands  $^6$
\and University of Maryland Baltimore County, 1000 Hilltop Circle, Baltimore, MD, 21250, USA 
\and  Jacobs University Bremen, Campus Ring 1, 28759 Bremen, Germany
\and Harvard-Smithsonian Center for Astrophysics, 60 Garden St., Cambridge, MA 02138, USA }

\date{Received, accepted }

\abstract{
We analyzed global properties, radial profiles, and 2D maps of the metal abundances and temperature in the cool core cluster of galaxies Hydra~A using a deep $\sim120$ ks XMM-Newton exposure. 
The best fit among the available spectral models is provided by a Gaussian distribution of the emission measure ({\it gdem}).
We can accurately determine abundances for 7 elements in the cluster core with EPIC (O, Si, S, Ar, Ca, Fe, Ni) and 3 elements (O, Ne, Fe) with RGS. 
The {\it{gdem}} model gives lower Fe abundances than a single-temperature model. Based on this, we explain why simulations show that the best-fit Fe abundance in clusters with intermediate temperatures is overestimated.
The abundance profiles for Fe, Si, S, but also O are centrally peaked. Combining the Hydra A results with 5 other clusters for which detailed chemical abundance studies are available, we find a significant decrease in O with radius, while the increase in the O/Fe ratio with radius is small within 0.1~$r_{200}$, where the O abundances can be accurately determined, with $d({\rm O/Fe})/d({\rm log}_{10}r/r_{200})=0.25\pm0.09$. 
We compare the observed abundance ratios with the mixing of various supernova type Ia and core-collapse yield models in different relative amounts.
Producing the estimated O, Si, and S peaks in Hydra~A requires either the amount of metals ejected by stellar winds to be 3--8 times higher than predicted by available models or the initial enrichment by core-collapse supernovae in the protocluster phase not to be as well mixed on large scales as previously thought.
The temperature map shows cooler gas extending in arm-like structures towards the north and south. These structures, and especially the northern one, appear to be richer in metals than the ambient medium and spatially correlated with the large-scale radio lobes. 
With different sets of assumptions, we estimate the mass of cool gas, which was probably uplifted by buoyant bubbles of relativistic plasma produced by the AGN, to $1.6-6.1\times10^9\:M_\odot$, and the energy associated with this uplift to $3.3-12.5\times10^{58}$ ergs. The best estimate of the mass of Fe uplifted together with the cool gas is $1.7\times10^7\:M_\odot$, 15\% of the total mass of Fe in the central 0.5\arcmin\ region. 
}
\maketitle

\footnotetext[6]{now a Chandra fellow at the Kavli Institute for Particle Astrophysics and Cosmology, Stanford University, 382 Via Pueblo Mall, Stanford, CA 94305-4060, USA}

\section{Introduction}

Clusters of galaxies provide a unique environment for elemental abundance measurements and for the study of the chemical enrichment history of the Universe, because their large potential wells retain all the metals  produced by the member galaxies. Of particular interest are clusters of galaxies showing a centrally peaked surface brightness distribution and a cool core, whose spectra are often richer in emission lines because of the lower central temperatures. In addition, these ``cool-core'' clusters have been shown to exhibit a central peak in the abundance distribution of several elements, in particular iron \citep[e.g.][]{DeGrandi01} and other metals produced by type Ia supernovae (SN~Ia), which led to the conclusion that the central excess is most probably due to enrichment by SN~Ia in the central dominant galaxies \citep[for a review, see][and references therein]{Werner08rev}. 

Cooling-core clusters are furthermore in the limelight because of the so-called ``cooling-flow problem'': the high surface brightness in the central peak implies a high density and a short cooling time, however the observed rate of cooling of the central gas is often orders of magnitude below what is expected in the absence of any heat sources \citep[for a review, see][]{Peterson06}. As a solution, it has been proposed that active galactic nuclei (AGN) in the central dominant galaxies can provide enough energy to the central gas to balance the cooling, since signs of energetic interaction between the AGN radio plasma and the intra-cluster medium (ICM) have been observed in many systems \citep[for a recent review, see][]{McNamara07}. Recently, the AGN-ICM interaction has also been shown, by theoretical models \citep{Rebusco06}, hydrodynamic simulations \citep{Roediger06} and observations \citep{simionescu2007b} to be a main mechanism for transporting the metals produced in the central galaxy into the ICM.

Hydra A was one of the first cooling-core clusters in which a displacement of X-ray gas in the center by radio lobes from the central AGN was found \citep{McNamara00} and it is still one of the most dramatic examples of AGN interaction \citep{David01, Nulsen02, Nulsen05}. In a deeper Chandra observation of the cluster, a sharp X-ray surface brightness edge was found at radii between 4.3 -- 6\arcmin\ (200 -- 300 kpc), interpreted as a shock wave caused by an AGN outburst with an estimated total energy of $10^{61}$ erg \citep{Nulsen05}, which is sufficient to balance radiative cooling for at least several $10^8$ yrs. Further features besides the shock and the inner cavities are a $\sim 60$ kpc long filament running from the inner cavities outwards, that may be reminiscent of the ``X-ray arms" in M87 \citep[e.g.][]{Feigelson87,Boehringer95,Belsole01,Forman06}, and surface brightness depressions in the outer region inside the shock front, which coincide with the outer radio lobes \citep{Wise07}. The relatively low temperature \citep[$T_X \sim 3.1-3.7$ keV,][]{David01} implies X-ray spectra rich in emission lines providing good spectroscopic diagnostics. 

The goal of the current work is twofold. Firstly, we aim to study the abundances, abundance ratios, and radial trends for different chemical elements. This can reveal clues about the chemical enrichment history of Hydra A and, by comparison with previous results from deep pointings of other nearby bright clusters, can broaden our overall understanding of the origin and distribution of metals in clusters of galaxies. Secondly, we investigate a two-dimensional metallicity map to determine what influence an AGN outburst as violent as the one in Hydra A can have on the distribution of chemical elements in the ICM.

Throughout the paper, we assume $H_0=70 {\rm \:km\: s^{-1}\: Mpc^{-1}}$, $\Omega_\Lambda=0.73$, and $\Omega_M=0.27$. At the redshift of Hydra A ($z=0.0538$), $1^{\prime\prime}$ corresponds to 1.05 kpc. Unless otherwise stated, the elemental abundances are given with respect to the proto-Solar values of \citet{lodders2003}, the errors are at the 1$\sigma$ level, and upper limits at the 2$\sigma$ level. 
The recent solar abundance determinations by \citet{lodders2003} give significantly lower abundances of oxygen, neon and iron than those measured by \citet{Angr}. Use of these new determinations affects only the units with respect to which we present the elemental abundances in our paper; the actual measured values can be reconstructed by multiplication of the given values with the solar normalizations. For a compilation of these normalizations for different commonly used abundance determinations, see the review of \citet{Werner08rev}.

\section{Observation and data analysis}

\subsection{EPIC analysis}

Hydra A was first observed with XMM-Newton on December 8th, 2000, for 32.6 kiloseconds (ks). A subsequent 123 ks observation was performed on May 11th, 2007. Since the second observation is significantly deeper and large parts of the first observation were affected by soft proton flares, we will focus in this work primarily on the second observation.  We extracted a lightcurve for each of the three detectors separately and excluded the time periods in the observation when the count rate deviated from the mean by more than 3$\sigma$ in order to remove flaring from soft protons \citep{Pratt02}. After this cleaning, the net effective exposure is $\sim$62 ks for pn, $\sim$81 ks for MOS1, and $\sim$ 85 ks for MOS2. We furthermore excluded CCD 5 of MOS2 from our analysis due to its anomalously high flux in the soft band during the observation \citep[see ][]{snowden2007}.
For data reduction we used the 7.1.0 version of the XMM-Newton Science Analysis System (SAS); the standard analysis methods using this software are described in e.g. \citet{Watson01}. 

For the background subtraction, we used a combination of blank-sky maps from which point sources have been excised \citep{ReadPonman,Carter07} and closed-filter observations. This is necessary because the instrumental background level of XMM-Newton is variable and increases with time. Both the blank-sky and the closed filter observations were transposed to a position in the sky corresponding to the orientation of XMM-Newton during our observation. We calculated the count rates in the hard energy band (10.--12. keV for MOS, 12.--14. keV for pn) outside of the field of view (OoFoV) for our observation, for the blank-sky maps and for the closed filter observations. These count rates are a good indicator for the level of instrumental background in each data set, since no photons from real X-ray sources should be recorded outside the field of view. For each detector we then added to the corresponding blank sky background set a fraction of the closed filter data designed to compensate for the difference between the OoFoV hard-band count rate in the observation and in the blank sky data. We note that simply scaling up the blank sky data to match the observed OoFoV hard-band count rate would implicitly also scale up the cosmic X-ray background (CXB) component contained in these blank sky maps, leading to an overestimation of the total background. We compared the blank sky background spectra of \citet{ReadPonman} and \citet{Carter07} with the current observation in a 10--14\arcmin\ annulus. We find a good agreement both in the soft (0.35--1 keV) and hard (5--10 keV) bands, while between 1 and 5 keV the spectrum from the observation shows an excess consistent with residual cluster emission with a temperature of around 2 keV. The best agreement with the 10--14\arcmin\ spectra, both in the soft and hard bands, is provided by the \citet{ReadPonman} background for the MOS detector and the \citet{Carter07} background for the pn detector. Consequently, we used these respective blank sky fields to estimate our background. 

Out-of-time events were subtracted from the PN data using the standard SAS prescription for the extended full frame mode.

\subsection{RGS analysis}

We extract the RGS spectra following the method described by \citet{tamura2001b}. 
We model the background using the standard background model available in SAS \citep[{\texttt{rgsbkgmodel}},][]{riestra2004}. 
The cluster spectra are extracted from a region which is 3\arcmin\ wide in the cross-dispersion direction of the instrument.
 
The line emission observed with the RGS from extended sources is broadened by the spatial extent of the source along the dispersion direction. 
In order to account for the line broadening in the spectral modelling, we convolve the line spread function (lsf) model with the surface brightness profile of the source along the dispersion direction. We derive the surface brightness profile from the EPIC/MOS1 image in the 0.8--1.4~keV band. Because the radial profile of an ion producing an observed spectral line can 
be different from the radial surface brightness profile in the broad band, we multiply the line profile with a scale parameter $s$, which is left as a free parameter in the spectral fit. 
The scale parameter $s$ is the ratio of the observed lsf width and the width of the lsf model convolved with the surface brightness profile. For a flat radial distribution of the line emitting ions, the scale parameter is $s=1$.    

\subsection{Spectral modeling}

We use the SPEX package \citep{kaastra1996} to model our spectra with a plasma model in collisional ionization equilibrium (MEKAL). We note that the MEKAL model implemented in SPEX has a more updated line catalog compared to that used by XSPEC; however, SPEX does not support any other plasma models, notably APEC. Several differential emission measure models which include contribution from gas with a range of different temperatures, rather than a single temperature approximation, are available in SPEX. Throughout the paper we will mainly use a Gaussian distribution of the emission measure ({\it gdem}) around the best-fit average temperature, of the form
\begin{equation}\label{eq:gdem}
\frac{{\rm d}Y}{{\rm d}x}=\frac{Y_0}{\sqrt{2\pi}\sigma_{\rm T}}e^{-\left(x-x_0\right)^2/2\sigma_{\rm T}^2} ,
\end{equation}
where $Y_0$ is the total, integrated emission measure ($Y_0 = \int n_{\mathrm{e}} n_{\mathrm{H}} dV$), $x\equiv{\rm log}T$ and  $x_0\equiv{\rm log}T_0$ with $T_0$ the average temperature of the plasma. Sect. \ref{gdemeff} gives a detailed explanation why this is the most appropriate available multi-temperature model for the data.

Unless otherwise stated, the Galactic absorption column density was fixed to $N_{\mathrm{H}}=4.8\times10^{20}$~cm$^{-2}$, the average value from the two available \ion{H}{i} surveys: the Leiden/Argentine/Bonn (LAB) Survey of Galactic \ion{H}{i} \citep[][ $N_{\mathrm{H}}=4.68\times10^{20}$~cm$^{-2}$]{kalberla2005} and the \ion{H}{i} data by \citet{dickey1990} ($N_{\mathrm{H}}=4.90\times10^{20}$~cm$^{-2}$). Point sources identified using the X-ray images are excluded from the spectral analysis. The spectra obtained by MOS1, MOS2 and pn are fitted simultaneously with their relative normalizations left as free parameters. To account for possible gain shifts, the redshift was left as a free parameter in the initial fit and fixed to its best-fit value when estimating the errors.

The global spectrum extracted from a circular extraction region with a radius of 3\arcmin\ (Sect. \ref{global}) is fitted in the 0.35--10 keV band. Because of the low number of counts and the high background above 7~keV, the spectra extracted from all other regions are fitted in the 0.35--7 keV band. The spectra extracted for the analysis of the global properties of the cluster and for the radial profiles are binned to the optimal bin size using the ``obin'' command of SPEX. This command rebins the data to at least 1/3 of the FWHM of the instrument depending on the count rate at the given energy. The spectra extracted for constructing 2D maps of the spectral properties of the cluster have much lower statistics and are consequently simply binned with a minimum of 30 counts per bin. 

We fitted the normalization, temperature and, where the statistics allowed, the abundances of O, Ne, Mg, Si, S, Ar, Ca, Fe, and Ni as free parameters. The abundances of the other elements heavier than He are fixed to 0.5 of the Solar value.

Due to the high photon statistics of this deep observation, our best fit reduced $\chi^{2}s$ are sensitive to calibration problems and to the differences between the individual EPIC detectors. To account for this, we include 3\% systematic errors over the entire energy band used for fitting the global properties of the cluster and the radial profiles. We usually obtain good ($<2\sigma$) agreement between the different detectors, the only notable exception is mentioned in Sect. \ref{res:radprof}.

In addition, to insure an optimal determination of element abundances, it is crucial to accurately fit the continuum around emission lines. This is not guaranteed by available emission measure model distributions, which are always a simplification of the true thermal structure in any given spectral extraction region. It is necessary therefore, after determining the best average temperature or temperature distribution using the full spectral band, to fix the thermal structure parameters and fit the elemental abundances in narrower bands around the emission lines of the respective elements, allowing the spectrum normalization to vary and thereby correct for small inaccuracies in the best determination of the continuum in those narrow energy bands. 
We accordingly fitted the Si, S, Ar, and Ca metallicities in the 1.5--5. keV energy band, and the O abundance in the 0.35--1.5 keV band. Unless otherwise stated, we report the best-fit Fe abundance using the full energy band (both Fe-L and Fe-K lines). 

The Ne abundance could not be constrained with the EPIC spectra because the \ion{Ne}{x} line lies in the Fe-L complex where the EPIC cameras cannot resolve the individual lines, thus a high Ne abundance can be confused with a higher contribution of cooler multi-temperature gas (a higher $\sigma_{\rm T}$) and vice versa. Consequently, the Ne abundance was fixed relative to Fe based on the RGS fit (Sect. \ref{section:RGS}). The Mg abundances could not be determined with more than 3$\sigma$ significance in any of the regions. Moreover, as reported by \citet{deplaa2007}, the systematic error in the effective area calibration of EPIC around the Mg energy is so large that the Mg abundance determination cannot be trusted. The Ni abundance was not determined in any of the regions for which the chosen spectral fitting band was 0.35--7 keV, which does not include the energy of the Ni lines. The abundances of Si, S and Fe, on the other hand, can be determined with good accuracy in most regions and, where the surface brightness is sufficiently high, we also obtain reliable O, Ar, Ca and Ni abundance determinations.

\section{Results}

\subsection{Global spectra}\label{global}

\begin{table}
\caption{Fit results for the EPIC data using a {\it gdem} model. $\sigma_{\rm T}$ is the width of the Gaussian temperature distribution. All abundances are relative to the proto-Solar units of \citet{lodders2003}.}
\label{tab:global}
\begin{center}
\begin{tabular}{l|cc}
\hline
\hline
Extraction region 			& 0--3\arcmin    	& 3--8\arcmin  \\
\hline
$Y$ 	($10^{66}$ cm$^{-3}$)			& $24.85\pm0.09$  	& $9.27\pm0.05$ \\
k$T$	 (keV)	  		& $3.42\pm0.01$   	& $3.65\pm0.03$ \\
$\sigma_{\mathrm{T}}$ (log$T$)	& $0.251\pm0.006$	& $0.210\pm0.017$\\
O/Fe                   		& $0.85\pm0.09$   	& -- \\  
Si/Fe                  		& $0.65\pm0.05$		& $0.45\pm0.07$ \\    
S/Fe                   		& $0.58\pm0.05$  	& $0.30\pm0.10$ \\     
Ar/Fe                  		& $0.48\pm0.14$   	& -- \\     
Ca/Fe                  		& $1.28\pm0.16$		& $0.47\pm0.33$ \\     
Fe                    		& $0.445\pm0.007$   	& $0.324\pm0.013$ \\      
Ni/Fe                    	& $1.35\pm0.20$   	& -- \\    
\hline 
$\chi^{2}$ / d.o.f. 		&  1449 / 1211    &  985 / 972  \\ 
\hline
\end{tabular}
\end{center}
\end{table}

\begin{figure}
\includegraphics[height=\columnwidth,angle=270,bb=74 34 582 724,clip]{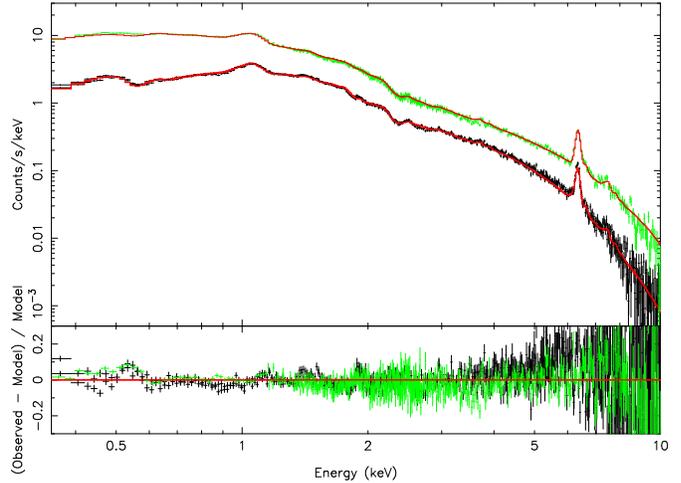}
\caption{The EPIC spectrum of the central 3\arcmin\ and fit residuals with respect to the best-fit Gaussian emission measure distribution model. Data points from pn are shown in green, MOS in black, and the best-fit model in red.}
\label{fig:epicsp}
\end{figure}

To characterise the global properties of the cluster and to determine accurate abundance values for as many elements as possible, we extract spectra from two large spatial regions: a circular region with a radius of 3\arcmin\ centered on the cluster core and an annulus with an inner radius of 3\arcmin\ and outer radius of 8\arcmin. This should insure good statistics in both regions, while keeping the number of source counts in the outer annulus well over the background level.

The best-fit width of the {\it gdem} distribution, $\sigma_{\rm T}$, as well as all other best-fit parameters are shown in Table~\ref{tab:global}. The $\sigma_{\rm T}$'s determined from our fits are significantly larger than zero both in the inner and outer annuli and the values are similar to those found by \citet{deplaa2006} in S\'ersic~159-03. A spectrum of the central 3\arcmin\ region and the best-fit model are shown in Fig. \ref{fig:epicsp}. It can be seen that the {\it gdem} model provides a good fit to both Fe-L and Fe-K lines, which a single temperature model did not achieve. For a more detailed discussion about the effects of the presence of such a multi-temperature structure on the spectral properties, see Sect. \ref{gdemeff}. 

The deep observation of Hydra~A in combination with the large collecting area of XMM-Newton make this one of the best cluster observations for investigating the chemical composition of the intra-cluster medium. The spectrum extracted from the circle with the radius of 3\arcmin\ allows us to determine relatively accurate abundance values for 7 elements including Ar and Ca which have relatively small equivalent widths (see Table~\ref{tab:global}). We note that there is a good agreement between all fit parameters determined using the MOS and pn spectra separately: the largest difference (2$\sigma$) is in determining the Ca abundance. Therefore the effect of calibration uncertainties should be small. The abundances in the cluster core relative to Fe are consistent with the values measured by \citet{deplaa2007} for a sample of 22 clusters. The absolute abundance values are lower in the outer part of the cluster than in the core. The observed relative abundances with respect to Fe are also lower in the outer part, although the differences are in most cases not significant.

We do not report the O/Fe value in the outskirts, which has a large error and is furthermore uncertain because of the low energy wing of the EPIC response which does not allow us to resolve the O line well and makes its abundance value strongly dependent on calibration uncertainties and on the uncertainties in the modelling of the low energy X-ray foreground, especially in regions of low surface brightness. The equivalent widths of the Ar and Ca are relatively low and the lower statistics in the outer extraction region make the abundance determination for these elements uncertain. Except for Si/Fe and S/Fe these uncertainties make it difficult to draw strong conclusions about the differences in the relative abundances between the inner and outer regions.

\subsection{Radial profiles}\label{res:radprof}

\begin{table*}[tb]
\begin{center}
\caption[]{Radial profiles obtained by fitting the EPIC data with a {\it gdem} model. The emission measure $Y$ was corrected for chip gaps and excluded point sources.}

\label{tab:profile}
\vspace{2mm}
\begin{footnotesize}
\begin{tabular}{l|ccccccc}
\hline
\hline
 & 0--0.5\arcmin & 0.5--1.0\arcmin & 1.0--2.0\arcmin & 2.0--3.0\arcmin & 3.0--4.0\arcmin & 4.0--6.0\arcmin & 6.0--8.0\arcmin \\
\hline
Y ($10^{66}$ cm$^{-3}$)      	& $5.15\pm0.03$ & $5.82\pm0.03$ & $7.25\pm0.03$ & $5.56\pm0.03$ & $3.66\pm0.02$ & $3.74\pm0.03$ & $1.89\pm0.02$ \\
$kT$ (keV)            	& $3.17\pm0.02$ & $3.44\pm0.02$ & $3.41\pm0.02$ & $3.57\pm0.03$ & $3.82\pm0.04$ & $3.60\pm0.04$ & $3.46\pm0.08$ \\ 
$\sigma_{\mathrm{T}}$ (log$T$) 	& $0.218\pm0.008$&$0.215\pm0.011$&$0.246\pm0.009$&$0.258\pm0.013$&$0.23\pm0.02$ & $0.23\pm0.02$ & $0.08\pm0.08$ \\
O/Fe             	& $0.80\pm0.09$ & $0.77\pm0.11$ & $0.85\pm0.13$ & $0.74\pm0.16$ & $0.59\pm0.24$ & $<0.31$       & $-$ \\ 
Si/Fe             	& $0.60\pm0.06$ & $0.67\pm0.06$ & $0.64\pm0.08$ & $0.81\pm0.13$ & $0.52\pm0.18$ & $0.63\pm0.16$ &  $0.80\pm0.33$ \\ 
S/Fe              	& $0.65\pm0.07$ & $0.56\pm0.08$ & $0.51\pm0.10$ & $0.74\pm0.16$ & $0.38\pm0.21$ & $0.53\pm0.22$ &  $<0.80$ \\ 
Ar/Fe             	& $0.85\pm0.18$ & $0.54\pm0.23$ & $0.62\pm0.26$ & $<0.57$       & $<0.69$       & $<0.63$       & $<0.65$ \\ 
Ca/Fe             	& $1.02\pm0.24$ & $1.50\pm0.27$ & $1.82\pm0.31$ & $1.52\pm0.49$ & $1.55\pm0.70$ & $<0.68$       & $ <0.72 $ \\ 
Fe                	&$0.550\pm0.012$&$0.483\pm0.013$&$0.392\pm0.010$&$0.308\pm0.012$&$0.287\pm0.016$&$0.323\pm0.018$& $0.29\pm0.07$ \\
$\chi^{2}$ / d.o.f	&  1388 / 1222  &  1211 / 1107  &  1196 / 1067  &  1025 / 1029   &   972 / 996  &   1063 / 1117  & 1071 / 917 \\ 
\hline

\end{tabular}
\end{footnotesize}
\end{center}
\end{table*}

\begin{figure}
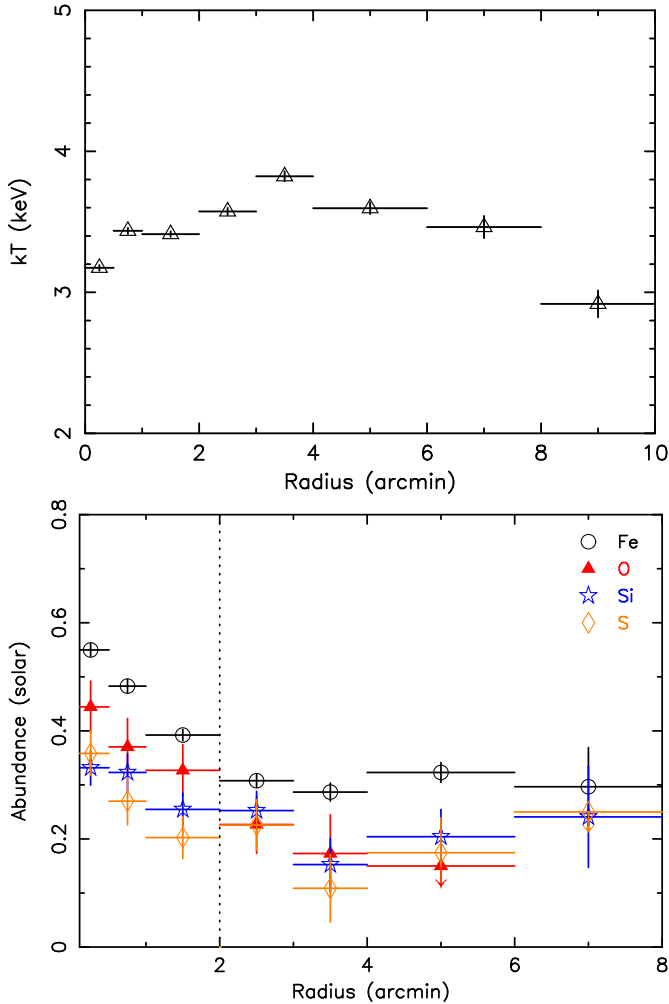

\includegraphics[height=\columnwidth,angle=270]{0225kT.ps}
\includegraphics[height=\columnwidth,angle=270]{0225rmet.ps}
\caption{{\it{Top panel: }}projected radial temperature profile of Hydra~A.   {\it{Bottom panel: }}radial profiles for four elements with the best determined abundance values. The radius at which the onset of an abundance peak is seen is indicated by a vertical dotted line. $r_{200}$ \citep[from][adapted to $H_0=70 {\rm \:km\: s^{-1}\: Mpc^{-1}}$]{reiprich2002} corresponds to 23.7\arcmin.}
\label{fig:profile}
\end{figure}

We also extracted spectra from 7 circular annuli centered on the cluster core in order to determine average radial temperature and abundance trends. Our results are summarized in Table \ref{tab:profile}.

The upper panel of Fig. \ref{fig:profile} shows that the radial temperature distribution has a dip in the core of the cluster and is relatively flat from 1\arcmin\ to 7\arcmin, with the exception of an elevated value in the 3--4\arcmin\ bin, which is likely associated with the large-scale shock in Hydra~A \citep[][]{Nulsen02, Nulsen05}. For a study of the thermal structure associated with the shock, see \citet{Simionescu08shock}. Beyond 7\arcmin, we see the  onset of a significant temperature decrease. 

The radial distribution of all investigated metal abundances peaks at the core of the cluster (see lower panel of Fig. \ref{fig:profile}). Radial gradients are seen in the distribution of elements which are predominantly produced both by SN~Ia (e.g. iron) and by core-collapse supernovae (\sncc, e.g oxygen). Again, the most robustly determined abundances are those for Fe, Si, and  S, but in the inner regions (within $\sim$3\arcmin) where the O abundance determination is less sensitive to background subtraction, its value is also quite accurate. 

In the last bin, uncertainties in determining the background at low energies become important. Thus, we do not report the O abundance, which is the most affected by such uncertainties, and present in Table \ref{tab:profile} the Fe abundance determined based on Fe-K only. The best-fit Fe abundance based on Fe-L and Fe-K combined would have been $0.43\pm0.03$ solar (consistent both for MOS and pn), which is unlikely at such a large radius from the cluster center.

\begin{figure}
\includegraphics[height=\columnwidth,angle=270]{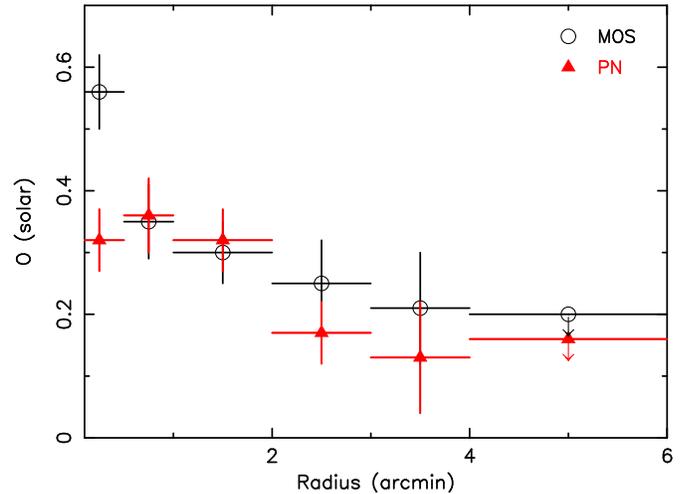}
\caption{Radial profiles for O obtained by fitting spectra from the MOS and pn detectors independently.}
\label{fig:omospn}
\end{figure}

To confirm the presence of a peak in the distribution of the O abundance which, as discussed above, is difficult to determine, we also checked the results from the MOS and pn detectors separately. The two obtained O abundance profiles are plotted in Fig. \ref{fig:omospn}. We find a discrepancy in the O determination from the two different detectors in the most central bin, but beyond this the agreement is very good and it is clear that in both detectors the O abundances in the inner 3 radial bins (central 2\arcmin) are systematically higher than outside the 2\arcmin\ radius, indicating that the peaked O distribution is indeed real. The discrepancy in the central bin might stem from calibration problems at this particular position of the detector, causing somewhat different values of $\sigma_{\rm T}$ (0.19 for MOS and 0.245 for pn) and different abundances. At other detector coordinates, the results agree well.

\citet{David01} obtain from Chandra data a temperature profile with a dip in the center down to 3.1~keV, increasing up to 3.8--4.0 keV at 3.4\arcmin, in very good agreement with what we find in Fig. \ref{fig:profile}. Our Fe profile is also roughly in agreement with that obtained from Chandra data, showing the most pronounced peak in the inner $\sim$90--150 kpc (1.5--2.5\arcmin). However, the XMM-Newton data gives much lower Si abundances, and does not confirm the very high Si/Fe ratio obtained previously by \citet{David01}. While the deprojection used for the Chandra analysis could account for relatively higher abundances of both Si and Fe compared to our projected radial profiles, it would not explain the discrepancy in the Si to Fe abundance ratios. 

\subsection{2D spectral properties}
\label{sect:2Dmaps}

\begin{figure}
\includegraphics[width=\columnwidth]{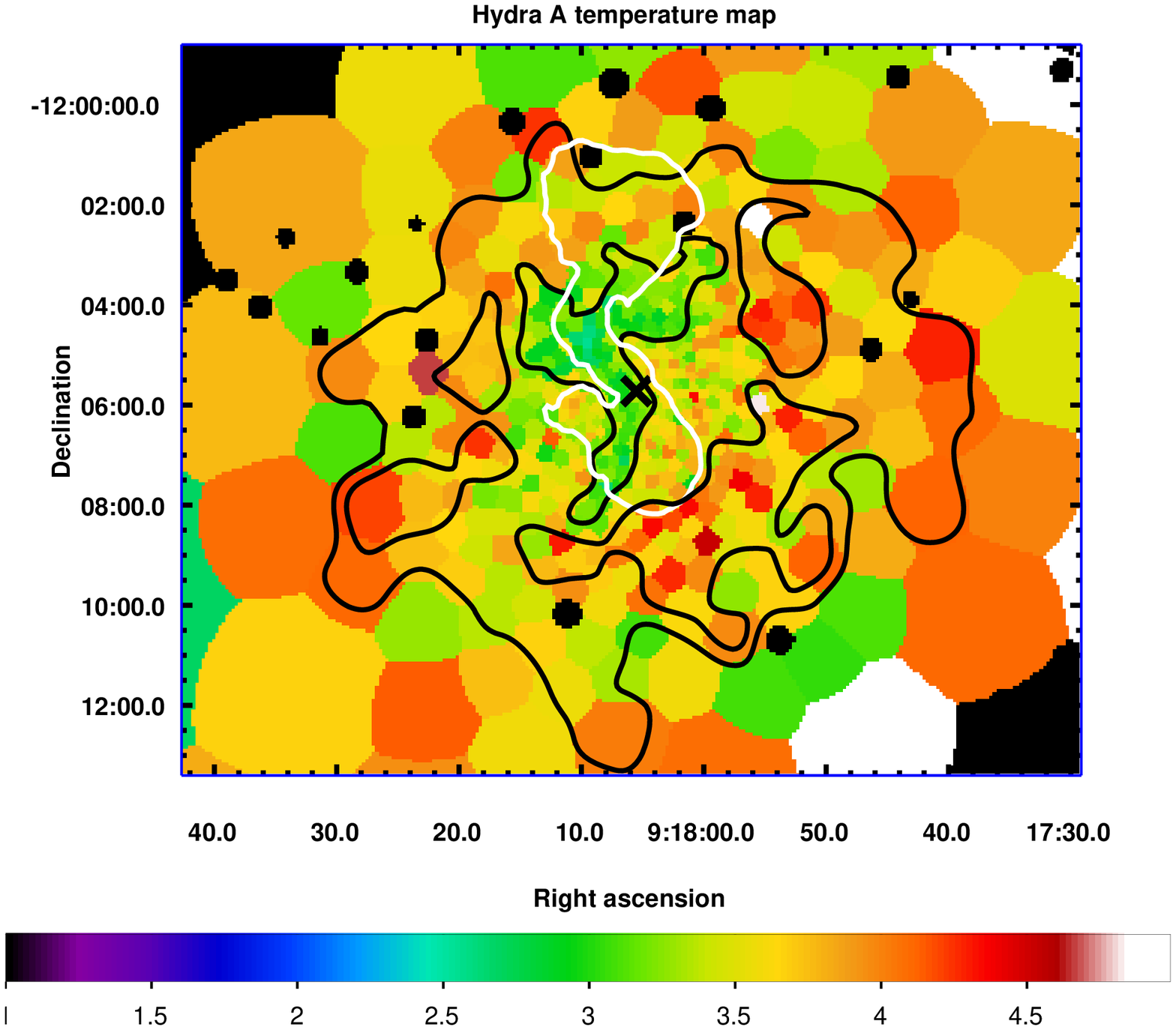}
\includegraphics[width=\columnwidth]{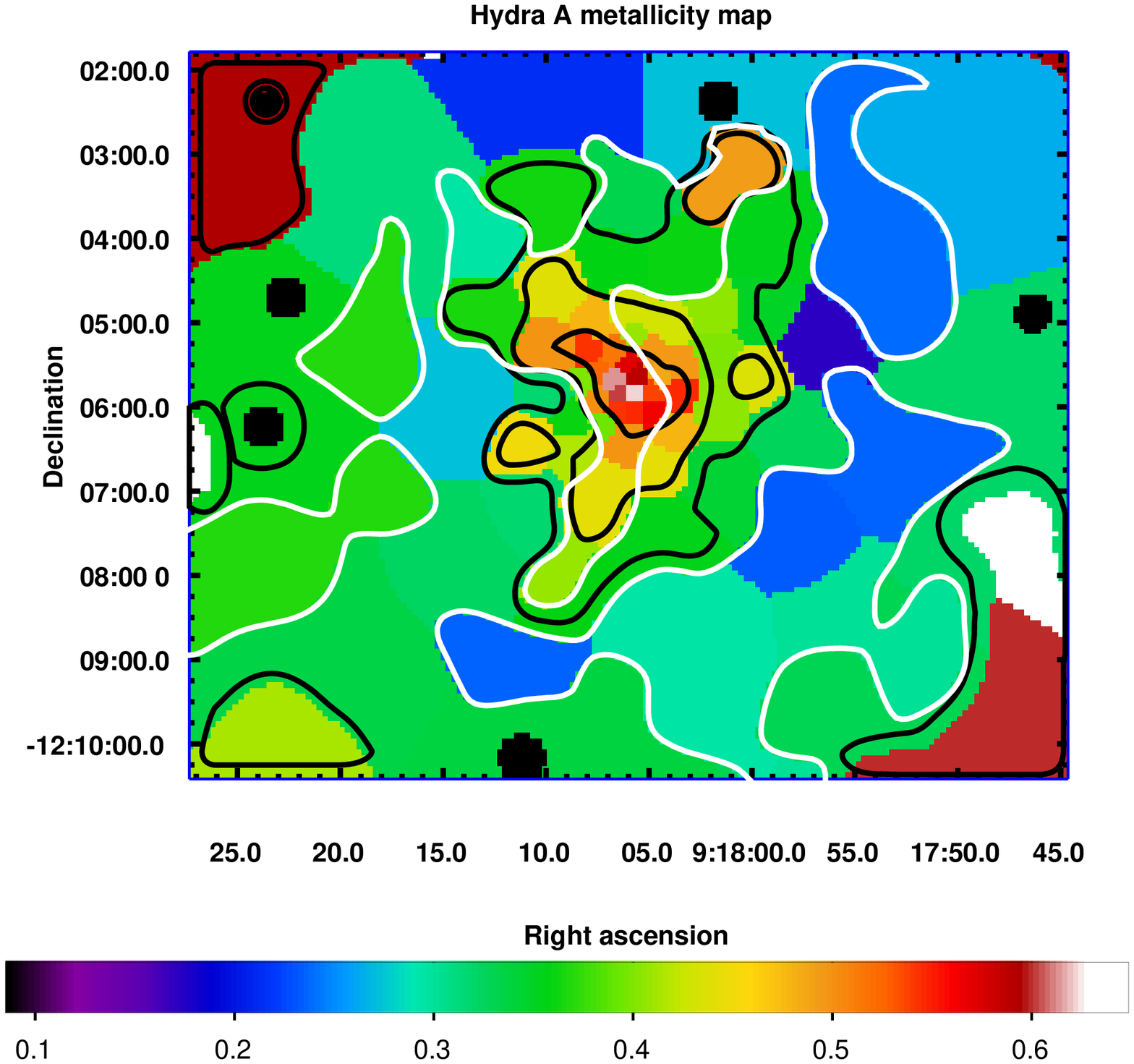}
\caption{{\it{Top panel: }} temperature map of the Hydra A cluster, using a minimum of $50^2$ counts per spatial bin (colorbar units are keV). Temperature contours are over-plotted in black, 327 MHz radio contours from \citet{Lane04} are over-plotted in white. The cross marks the surface brightness peak associated with the cluster center. {\it{Bottom panel: }} Fe abundance map, using a minimum of $150^2$ counts per spatial bin. Contours are over-plotted in black \citep[colorbar units are solar,][]{lodders2003}, temperature contours are over-plotted in white. Note that the Fe map is a zoom-in of the region shown in the temperature map.}
\label{fig:tmap}
\end{figure}

To produce two-dimensional maps of the temperature and abundance distribution in the cluster, we divided our observation into spatial bins with a fixed minimum number of counts employing an adaptive binning method based on weighted Voronoi tessellations \citep{Voronoi06}, which is a generalization of the algorithm presented in \cite{Cappellari03}. The advantage of this algorithm is that it produces smoothly varying bin shapes that are geometrically unbiased and do not introduce artificially-looking structures. We created a background-subtracted count map of the observation in the energy range 0.4-7.0 keV, combining all three EPIC detectors, and binned this map to $50^2$ counts per bin for generating the temperature map and $150^2$ counts per bin for generating an abundance map, which requires higher statistics. The binning to $50^2$ counts was constrained to follow the large-scale radio lobes seen at 327 MHz \citep[][see white contours in the upper panel of Fig. \ref{fig:tmap}]{Lane04}, while the $150^2$-count bins were constrained to follow the temperature contours shown in black in the upper panel of Fig. \ref{fig:tmap}. We furthermore constrained the bins in the central regions (with a high count rate) not to be smaller than the extent of the XMM point-spread function (PSF). Since the statistics do not allow a reliable determination of any abundance except that for Fe, we fixed the O, Ne, Mg, Si, S, Ar, Ca and Ni abundances in the fit relative to Fe by the ratios representative of the global properties in the inner 3\arcmin, as presented in Sect. \ref{global}. Also, in the bins with $50^2$ counts, the statistics do not allow us to fit the spectra with free $\sigma_{\rm T}$. We therefore fixed it to 0.2, a typical value found in the radial profiles in Sect. \ref{res:radprof}. We note that fitting the bins for the temperature map with a single temperature model, both with free and fixed $N_{\mathrm{H}}$, shows the same structures. In the bins with $150^2$ counts used to determine the metallicity map, we can obtain best-fit values for $\sigma_{\rm T}$, however these are very uncertain, resulting in a very noisy $\sigma_{\rm T}$ map, and almost always in agreement with 0.2 (only one bin out of 80 is more than 5$\sigma$ away from 0.2 and three bins more than 2.5$\sigma$). The obtained metallicity maps with $\sigma_{\rm T}$ free and fixed to 0.2 are very similar and show the same features. We therefore choose to present the Fe abundance map obtained using $\sigma_{\rm T}=0.2$ in the lower panel of Fig. \ref{fig:tmap}. 

The temperature map clearly shows cooler gas extending in arm-like structures towards the north and south. The cool gas structure to the north follows well the northern radio lobe, which rises from the cluster core towards the NE and then either bends towards the NW (however, the bend in the cool gas feature is stronger than that of the radio lobe) or extends out into an umbrella/mushroom-shape at the position where the N radio lobe bends. The cool gas towards the south is also most probably associated with the AGN activity, although the feature in the temperature map is oriented at a slightly more easterly angle than the southern radio lobe. The Fe abundance map also shows features extending towards the north and south, in the same direction as the cool gas features, suggesting that the cool gas is more abundant than the surrounding halo. The cool ``arm'' to the NE is associated with a bright 1\arcmin\ long filament found by \citet{Nulsen05} using Chandra data.

\subsection{High resolution spectra}
\label{section:RGS}

\begin{figure}
\includegraphics[height=\columnwidth,angle=270]{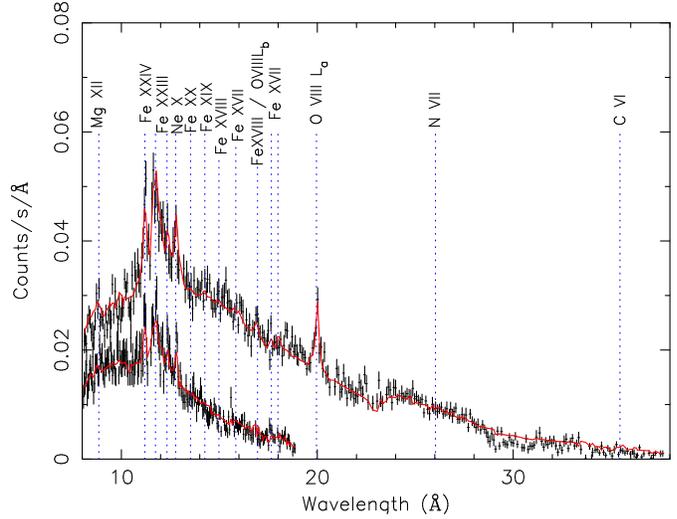}
\caption{Combined RGS $~1^{st}$ and 2$^{nd}$ order spectrum of Hydra~A extracted from a 3\arcmin\ wide strip 
centered on the core of the cluster. The continuous line represents the fitted model. 
The \ion{Fe}{xvii}--\ion{Fe}{xx} lines emitted by cooling gas between the \ion{Ne}{x} and \ion{O}{viii} lines are not visible in the spectrum. }
\label{fig:rgs}
\end{figure}

\begin{table}
\caption{RGS fit results using a 3\arcmin\ wide extraction region centered on the cluster core.}
\begin{center}
\begin{tabular}{l|cc}
\hline
\hline
 			& 2T    	& {\it gdem}+1T  \\
\hline
$Y_1$ ($10^{66}$~cm$^{-3}$)		& $21.58\pm0.22$   &   $21.29\pm0.25$    \\
k$T_1$ (keV)	& $2.37\pm0.06$	& 3.42 (fixed)\\
$\sigma_{\rm T1}$ (log$T$) & 0.0 (fixed) & $0.19\pm0.03$\\
$Y_2$ ($10^{66}$~cm$^{-3}$)		& $0.30\pm0.06$	& $0.26\pm0.04$	\\
k$T_2$ (keV)    & $0.64\pm0.04$	& $0.62\pm0.04$	\\
O/Fe 		& $0.74\pm0.10$	 &   $0.76\pm0.11$   \\
Ne/Fe		& $0.73\pm0.18$	 &    $0.84\pm0.20$	\\
Fe		& $0.23\pm0.02$	     	& $0.35\pm0.03$\\
Scale $s$ 	& $0.95\pm0.14$	  &  $1.01\pm0.20$   \\
\hline
$\chi^2$ / d.o.f. & 1145 / 915		    & 1273 / 915 \\
\hline
\end{tabular}
\label{tab:rgs_180}
\end{center}
\end{table} 

The spectrum from an extraction region which is 3\arcmin\ wide in the cross dispersion direction and effectively $\sim$10\arcmin\ long in the dispersion direction allows us to determine more accurate projected O/Fe and Ne/Fe abundances than possible with EPIC. RGS also allows us to put better constraints on the presence of cooling gas in the cluster core than EPIC by looking at individual lines of the Fe-L complex. 

The projected global temperature determined by RGS is $\sim$1~keV lower than that determined by EPIC. Similar differences between the mean temperatures determined by EPIC and RGS were found in other clusters \citep[e.g. 2A 0335+096, M~87;][]{werner2006,werner2006b}. In the case of Hydra~A, the reason for this discrepancy could be the fact that the RGS fit is performed in the soft-energy band, which is not sensitive to emission from higher-temperature gas present in the {\it gdem} model (see Sect. \ref{dem} for a more detailed discussion). To check this, we also performed a fit to the RGS data fixing the temperature to the best-fit EPIC temperature in the central 3\arcmin\ and leaving the $\sigma_{\rm T}$ free. We obtain a value for $\sigma_{\rm T}$ comparable to the typical results from EPIC. The elemental abundances in this case are considerably higher than for the single-temperature case, and thus in better agreement with the absolute values determined with EPIC. The abundance ratios, on the other hand, are unchanged. The reduced $\chi^2$ is somewhat worse for the {\it gdem} fit than for the single temperature fit, which is partly due to the low-temperature wing of the {\it gdem} approximation not being a perfect description of the temperature structure and partly due to the fact that the RGS/EPIC cross-calibration is not fine-tuned enough and the RGS and EPIC spatial extraction regions are different, such that fixing the temperature to the best-fit EPIC temperature causes a poorer fit.

While in several other cooling cores observed with RGS \citep[M~87, 2A~0335+096, Centaurus;][]{werner2006b,werner2006,sanders2008}, the spectral lines from the intermediate ionization states of Fe (\ion{Fe}{xvii}, \ion{Fe}{xviii}, \ion{Fe}{xix}, \ion{Fe}{xx}) are well resolved, we do not observe these lines in Hydra A. However, the spectrum fitted with a single temperature model or with a {\it gdem} model leaves strong residuals between $\sim$13~\AA\ and $\sim$19~\AA.
These residuals can be well fitted, both for the single-temperature and {\it gdem} models, by adding a cool gas component with k$T\approx0.6$~keV and emission measure of $\approx$1\% of the emission measure of the hot gas. The presence of this cool gas is detected with 5.0 and 6.5$\sigma$ significance for the single-temperature and {\it gdem} models, respectively. The fit results shown in Table~\ref{tab:rgs_180} and Fig.~\ref{fig:rgs} assume that the abundances of the cooler gas are the same as the abundances of the hot gas. 

The best fit O/Fe ratio is consistent with the value determined by EPIC. At the 3~keV temperature of Hydra~A, carbon and nitrogen are almost completely ionised and their lines cannot be detected by the RGS.

\section{Discussion}

\subsection{Spectral effects of the multi-temperature structure}\label{gdemeff}

Throughout this paper, we used a Gaussian distribution of the emission measure around the best-fit average temperature to model our spectra. We will explain in this section why this is necessary, what the implications on the best-fit spectral parameters are, and we will evaluate how accurate the {\it gdem} approximation is.

Usually, the strongest evidence of the presence of multi-temperature structure is the shape of the Fe-L complex, which is located at an energy where the effective area of the detectors is high, and whose shape changes drastically with the temperature of the emitting gas, especially for temperatures below $\sim$2 keV \citep[e.g.][]{BoehringerFeL}. 
In Hydra A, however, the shape of the Fe-L complex in the EPIC spectrum does not show strong indications of the presence of cooler gas as observed in other cooling core clusters \citep[e.g.][]{werner2006,deplaa2006,simionescu2007b}. 
The evidence of a multi-temperature structure in this case lies primarily in the fact that, when fitting spectra with high statistics, it becomes evident that the best-fit temperatures obtained fitting independently the soft (0.35--2 keV) and hard (2--7 keV) energy bands are in disagreement. Moreover, when using the entire energy band, the spectral model cannot simultaneously fit both Fe-L and Fe-K lines adequately. In particular, the Fe-K line in the data shows a shoulder towards higher energies, indicating Fe-K$\beta$ emission which suggests the presence of gas at higher temperatures than the fitted value from the single temperature model.

\begin{figure}
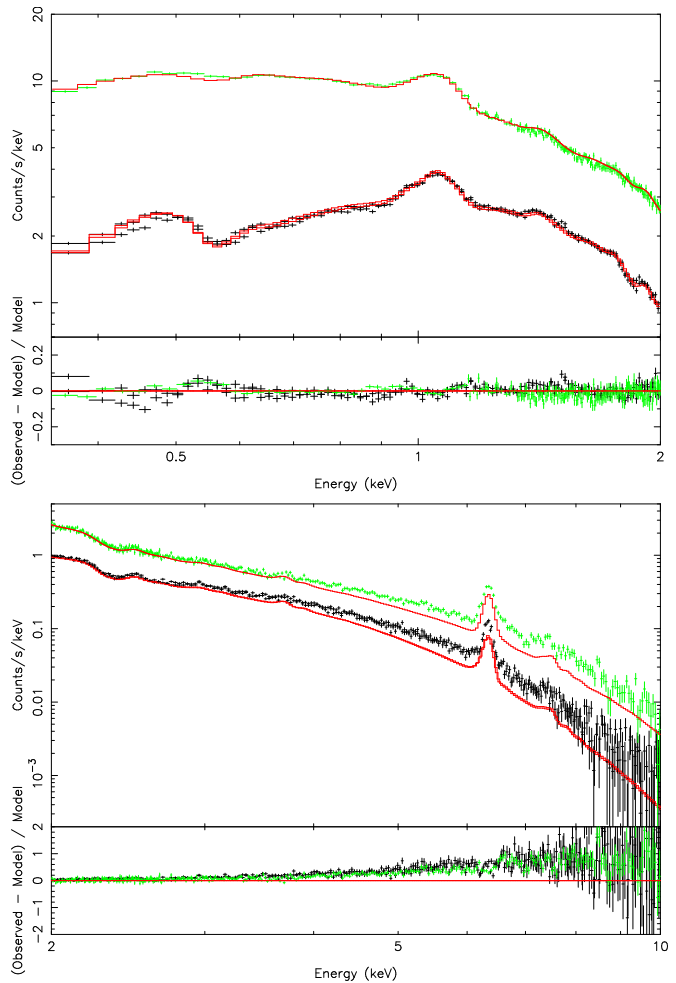

\includegraphics[height=\columnwidth,angle=270,bb=74 34 582 724,clip]{0225soft.ps}
\includegraphics[height=\columnwidth,angle=270,bb=74 34 582 724,clip]{0225sbex.ps}
\caption{{\it{Top panel: }}. EPIC spectrum from the central 3\arcmin\ region in the soft band and best-fit single temperature model.  {\it{Bottom panel: }} EPIC spectrum from the central 3\arcmin\ region in the hard band and the extrapolation of the best-fit single temperature model from the soft band. Data points from pn are shown in green, MOS in black, and the best-fit model in red. The clear residuals show the need for a multi-temperature model.}
\label{softhard}
\end{figure}

In Fig. \ref{softhard} we illustrate the difference between the soft and hard-band fits for the central 3\arcmin\ spectrum. The upper panel shows the best-fit single-temperature model in the soft band, which provides a very good fit to the Fe-L complex indicating the lack of a significant amount of cooler gas. The lower panel shows this model extrapolated to the hard band. The clear discrepancy indicates the fact that the best-fit single temperature model from the soft band does not account for the presence of gas at higher temperatures present in the spectrum. 
For the central 3\arcmin\ spectrum, the best-fit soft-band temperature is $2.81\pm0.05$ keV, while the hard-band temperature is $3.86\pm0.03$ keV, approximately the values spanned by the temperature profile and the 2D maps (Sect.~\ref{res:radprof} and ~\ref{sect:2Dmaps}).
A Gaussian emission measure distribution is one of the simplest models which can account for such a mixing of emission from plasma spanning a wider range of temperatures. 
Using this model significantly improves the fit ($\chi^2$/d.o.f. 1449/1211) with respect to a single temperature model ($\chi^2$/d.o.f. 2005/1212).

\subsubsection{Intrinsic multi-temperature structure}

\begin{figure}
\includegraphics[height=\columnwidth,angle=270]{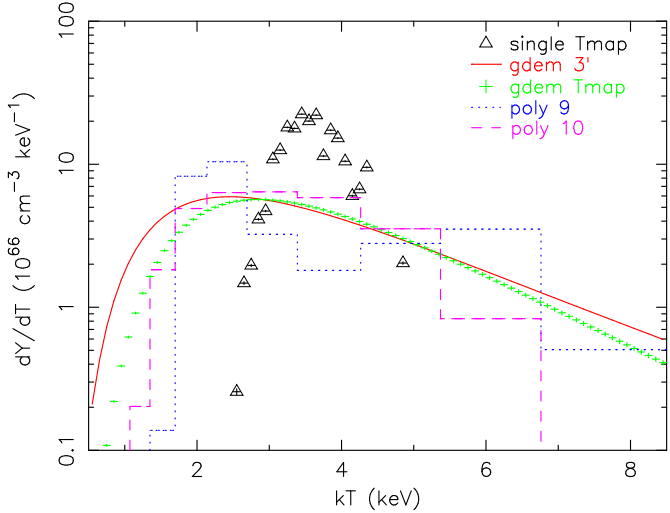}
\caption{Total $dY/dT$ curve constructed from the bins in the central 3\arcmin\ of the temperature map, showing the multi-temperature structure: in black, assuming single-phase in each bin; in green, assuming intrinsic {\it gdem} emission measure distribution with $\sigma_{\rm T}=0.2$ in each bin. The red curve shows the best-fit {\it gdem} model for the 3\arcmin\ region (Table \ref{tab:global}). In blue and magenta, different best-fit polynomial {\it dem} models for the 3\arcmin\ spectrum (Sect. \ref{dem}).}
\label{fig:histo}
\end{figure}

The multi-temperature structure can be partly due to the mixing of different regions with slightly different temperatures in the 2D map and partly due to intrinsic multi-temperature structure in each of these regions. To disentangle these two effects, we took the average temperatures and the emission measures in all the Voronoi bins with $50^2$ counts within a radius of 3\arcmin\ and, assuming the gas in each spatial bin to be single-phase, constructed a total $dY/dT$ curve, shown in Fig \ref{fig:histo}. We over-plot also the best-fit {\it gdem} model for the central 3\arcmin\ spectrum. As the figure shows, the shape of the $dY/dT$ curve due solely to spatial variations in the different bins does also have a Gaussian shape, however it is much narrower than the best-fit value from the {\it gdem} model (full-width at half maximum of about 1 keV compared to the 4.2 keV corresponding to $\sigma_{\rm T}=0.251$). With the statistical errors on $\sigma_{\rm T}$ from Table \ref{tab:global}, the difference is highly significant ($\approx35\sigma$). Note that hydrodynamic simulations of clusters which have recently undergone energetic events show a similarly wide $dY/dT$ curve \citep{Rasia06} as our best-fit {\it gdem} model.

Assuming, instead of isothermality, a {\it gdem} distribution with $\sigma_{\rm T}=0.2$ in each spatial region, we find a significantly better agreement between the total $dY/dT$ curve constructed from the bins in the temperature map and the best-fit {\it gdem} model for the central 3\arcmin\ spectrum. Remaining small disagreements could be explained for example by deviations of  $\sigma_{\rm T}$ from 0.2 in different regions. However, Fig \ref{fig:histo} does clearly suggest the need for intrinsic multi-temperature structure in each spatial bin. 

According to the best-fit {\it gdem} model, roughly 9.5\% of the total emission measure in the central 3\arcmin\ comes from gas between 0.5--1.5~keV, 10.7\% from gas between 1.5--2.0~keV, and 9.2\% from gas between 6--8~keV. In part, this intrinsic multi-temperature structure can be due to projection: on one hand, the large-scale shock in Hydra~A can produce the hard-tail in the $dY/dT$ distribution \citep[for more quantitative analysis, see][]{Simionescu08shock}; on the other hand, cool gas in the cluster outskirts can add to the tail at low temperatures. The low density in the outskirts makes it however unlikely that all the cool gas arises from projection alone and suggests the possible presence of denser, unresolved cool gas blobs. To make this statement more quantitative: the observed temperature profile in Hydra~A peaks at roughly 3.8 keV (in agreement with the deprojected profile of \citet{David01} from Chandra data). An average temperature of 2.0 keV thus is a factor 0.53 less than the peak temperature. If we compare for example with the scaled temperature profiles of \citet{vikhlinin2006}, we do not expect the temperature in the outskirts to drop to 0.53 of the peak temperature until around $r_{500}$, where little cluster emission is expected. Since we consider projection along a cyllinder of 3\arcmin\ radius, the volume does not increase significantly at large radii ($dV$ converges to $\pi(3^\prime)^2dr$ for large $r$ and is larger for smaller $r$ because of the curvature of the spherical shell with radius $r$ contained inside the considered cyllinder). Thus, unless we have reason to believe in a very small $dT/dr$ in the outskirts compared to the cluster center, which scaled temperature profiles do not suggest, $dY/dT =  n_{\mathrm{e}} n_{\mathrm{H}} dV/dT \propto  n_{\mathrm{e}}^2 dr/dT$ should decrease roughly as the square of the electron density beyond $\sim$3\arcmin, where projection effects come into play. According to \citet{vikhlinin2006}, the density drops by a factor of 15 between 0.23$r_{500}$ (3\arcmin) and $r_{500}$, thus the emission measure should decrease by at least a factor of $\sim$200. Clearly, Fig. \ref{fig:histo} shows that the emission measure around 2~keV is only a factor of a few, as opposed to a factor of a few hundred, below the emission measure from temperatures expected around 3\arcmin\ radius.
This shows that projection cannot have an important contribution and that the observed cool gas in the multi-temperature distribution must come from unresolved cool gas blobs in the central parts.

\subsubsection{Comparison with other available multi-temperature models}\label{dem}

The {\it gdem} model provides a significant improvement over a single temperature model, yet the real temperature distribution is probably even more complex, therefore we should test the reliability of the Gaussian approximation. 

Unfortunately, other available models except single-temperature and {\it gdem} are either poorly constrained by the current data set or give unphysical results. A power-law shaped emission measure distribution ({\it wdem}), which provides a good fit to other cooling-core cluster spectra \citep[e.g.][]{werner2006,deplaa2006,simionescu2007b}, requires a very flat, poorly constrained slope, which is what would be expected if the best approximation of the emission measure is Gaussian and we try to approximate it with a power-law. 

A two-temperature fit to the central 3\arcmin\ spectrum requires k$T_1=2.05^{+0.04}_{-0.05}$ keV and k$T_2=5.08^{+0.20}_{-0.13}$ keV with normalizations $Y_1=9.8^{+0.8}_{-0.5}$ and $Y_2=11.7^{+0.4}_{-0.5} \times10^{66}$ cm$^{-3}$. This fit gives only a small improvement with respect to the {\it {gdem}} fit ($\chi^2$/d.o.f.=1379/1209 compared to 1449/1211). The best-fit k$T_1$ and k$T_2$ are approximately the minimum and maximum temperatures represented in the {\it gdem} model within 1$\sigma_{\rm T}$ and the normalizations of the two components are similar, which is again what we would expect if the best approximation of the emission measure is Gaussian and we try to fit it with a two-temperature model (in the two-temperature model, the contribution of any phase between the two best-fit temperatures can always be reproduced by combining these two phases with appropriate normalization ratios). The {\it gdem} model thus provides an equally good fit and a more physical interpretation of the spectrum than the two temperature model, because it is difficult to imagine a scenario which would generate $\sim2$ and $\sim5$ keV gas in roughly equal amounts, no gas at intermediate temperatures and, implicitly, no mixing between the two gas phases. If the 5 keV component comes from the projected shocked gas layer, for instance, which would fulfill the requirement for the absence of mixing, it should certainly contain much less gas than the entire cluster center which would constitute the cooler phase.

The last test is to fit a generic differential emission measure model ({\it dem}) available in SPEX. 
For a grid of temperatures spaced by 0.1~log$T$ between a user-defined minimum and maximum value, this model constructs a spectrum corresponding to each grid point, and then can apply several methods to determine how to optimally combine these spectra in order to reproduce the data \citep[for further details, see][]{kaastra1996}. We chose the polynomial method, which finds the best $dY/dT$ in the form of a polynomial of a desired degree. Other methods gave either very similar results to the polynomial method or were computationally unstable. 
The disadvantage of the {\it dem} model is that it cannot fit any other parameters except those related to the $dY/dT$ behavior. We thus had to fix the elemental abundances to the best-fit {\it gdem} values. 

We show in Fig. \ref{fig:histo} the best-fit $dY/dT$ curves obtained from the {\it dem} model using $9^{\rm th}$ and $10^{\rm th}$ degree polynomials. It is immediately clear that, also using this method, we find a very broad multi-temperature distribution, similar to that required by the Gaussian model. The $10^{\rm th}$ degree polynomial agrees very well with the {\it gdem} model between $\sim$2--5~keV and falls somewhat faster at lower and higher temperatures. The $9^{\rm th}$ degree polynomial agrees with the {\it gdem} model at large temperatures but shows a hint of a bimodal $dY/dT$ distribution with peaks at 2.5 and 6 keV (however, still with significant amounts of emission between 2.5 and 6.0 keV, excluding a simple 2T approximation). While keeping in mind these small differences, we can conclude that the broad best-fit Gaussian emission measure distribution is a good simple approximation to the best-fit results of the polynomial {\it dem} model. 

It is interesting to further note the fact that the $dY/dT$ curves for both the {\it gdem} fit to the central 3\arcmin\ spectrum and for the $9^{\rm th}$ and $10^{\rm th}$ degree polynomial {\it dem} models peak around 2.3--2.5 keV, the best-fit temperature from RGS. This suggests that the discrepancy between the best-fit RGS and EPIC temperatures is partly due to the more limited RGS band, which corresponds to low energies and is therefore insensitive to the presence of the hard tail in the real temperature distribution.

\subsubsection{Confusion with soft excess}

Another way to account for temperature differences in the soft- and hard-bands (Fig. \ref{softhard}) is to allow the Galactic absorption column density to vary in the spectral fits. A best-fit $N_{\mathrm{H}}$ lower than the real value, allowing for emission from gas at low energies, in combination with a best-fit temperature slightly higher than the real value, can combine to give a good description of both the soft and hard part of the spectrum. If the multi-temperature structure is such as we describe above, the single-temperature fit with free $N_{\mathrm{H}}$ will be a significant improvement compared to the single-temperature fit with fixed $N_{\mathrm{H}}$. 

\begin{figure}
\includegraphics[height=\columnwidth,angle=270]{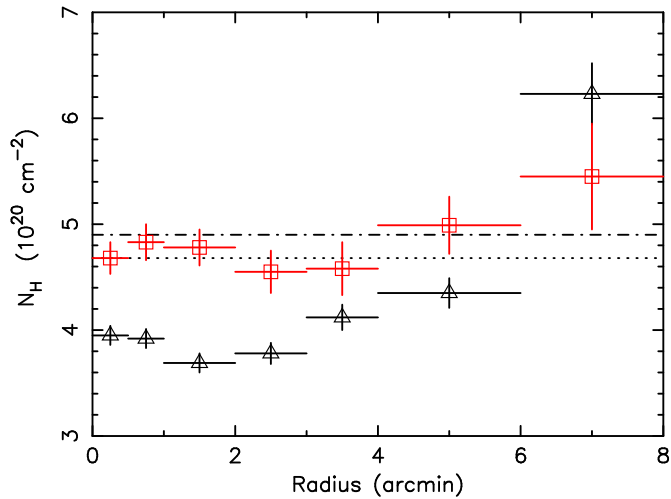}
\caption{ $N_{\mathrm{H}}$ profile determined fitting a single-temperature model to the full (0.35--7 keV, black triangles) and soft (0.35--2 keV, red squares) energy bands. The values determined from \ion{H}{i} surveys \citep{kalberla2005,dickey1990}. are over-plotted with a horizontal dotted and dash-dotted line, respectively.}
\label{fig:nh}
\end{figure}

We show in Fig. \ref{fig:nh} the $N_{\mathrm{H}}$ profile determined fitting a single temperature model to the full (0.35--7 keV) energy band. The $N_{\mathrm{H}}$ has a dip on the cluster center, which can be mistaken for the possible presence of soft excess generated either by non-thermal inverse Compton scattering of cosmic-microwave photons on relativistic electrons in the cluster or by the presence of a soft ($\sim0.2$ keV) thermal component \citep[for a review on soft excess, see][]{Durret08}. However, the $N_{\mathrm{H}}$ profile determined fitting a single temperature model only to the soft (0.35--2 keV) energy band is flat and in good agreement with the value determined from \ion{H}{i} surveys, disproving this scenario.

The Gaussian emission measure distribution model with $N_{\mathrm{H}}=4.8\times10^{20}$~cm$^{-2}$ provides a better fit than a single temperature model with free $N_{\mathrm{H}}$ ($\chi^2$/d.o.f. 1449/1211 compared to 1718/1211 for the central 3\arcmin\ region). Fitting a {\it gdem} model with free $N_{\mathrm{H}}$ further improves the fit to $\chi^2$/d.o.f.=1291/1210, with a best-fit $N_{\mathrm{H}}=3.92\pm0.07\times10^{20}$~cm$^{-2}$, still significantly lower than the Galactic value but closer to it than the best-fit $N_{\mathrm{H}}$ using a single-temperature fit ($3.69^{+0.06}_{-0.11}\times10^{20}$~cm$^{-2}$). Since the best-fit $N_{\mathrm{H}}$ using only the soft band is in agreement with the Galactic value ($4.74\pm0.07\times10^{20}$~cm$^{-2}$), the lower $N_{\mathrm{H}}$ in the full band must reflect some residual fitting problems. The most important effect is that the fitting tries to compensate for the residuals around the O-edge (around 0.5 keV, see Fig. \ref{fig:epicsp}) while constraining the average temperature by the continuum at higher energies; on the other hand, there might be some uncertainties in the calibration of the spectral slope of the EPIC detectors at low/high energies. Alternatively, the low $N_{\mathrm{H}}$ could be due to the presence of more cool gas in the real multi-temperature distribution than accounted for by the best-fit {\it gdem} model, although Sect. \ref{dem} suggests that this is not the case.

\subsubsection{The ``inverse'' Fe bias}

The best-fit Fe abundances from fitting the soft- and hard-bands individually with single-temperature models are both significantly lower than the best-fit Fe abundance determined from a single-temperature fit of the full spectral band (for the central 3\arcmin\ spectrum, $0.37\pm0.02$ solar for the soft band, $0.41\pm0.01$ for the hard band, $0.499\pm0.007$ for the full band). The discrepancy with the best-fit Fe abundance from the Gaussian emission measure distribution model ($0.445\pm0.007$) is smaller. This can be explained in the following way. The strength of the Fe-L complex is, for the same Fe abundance, higher for a cooler temperature. Similarly, the strength of the Fe-K complex is higher for a hotter temperature. If the model allows for some contribution to the spectrum of cooler and hotter gas compared to the average, a lower Fe abundance can reproduce the same Fe-L and Fe-K strengths for which a larger Fe-abundance would be needed if only gas at the average temperature is allowed in the model. This is especially important in clusters with intermediate average temperatures (2--4 keV), where both Fe-L and Fe-K emission lines are seen with relatively similar statistics, thus neither of the two lines predominantly drives the fit of the Fe abundance.

This effect is opposite to that described by \citet{buote1998}, in which not considering the presence of cooler gas in the multi-temperature structure leads to an underestimation of the real Fe abundance. That effect occurs primarily in cool systems (with an average k$T$ below about 2 keV) where the Fe abundance is determined largely based on the Fe-L complex, since Fe-K emission at those temperatures is very weak.

In our case, on the other hand, neglecting the contribution of both cooler {\it and} hotter gas in a Gaussian-like emission measure distribution leads to an overestimation of the Fe metallicity if both Fe-L and Fe-K lines are seen with good statistics in the spectra. We are able thus to explain exactly the bias found by \citet{Rasia08}, who analyzed six simulated galaxy clusters processed through an X-ray Map Simulator (X-MAS), which allowed to create mock MOS1 and MOS2 observations. They showed that the Fe abundances determined from fitting the mock X-ray spectra is systematically overestimated for intermediate (2--3 keV) clusters, while for both hot and cold systems, where either only the Fe-L or only the Fe-K emission lines dominate, the Fe abundance is recovered with good accuracy.

\subsection{Radial distribution of the chemical elements and comparison with other clusters}\label{disc:rad}

\begin{figure}[!h]
\includegraphics[height=\columnwidth,angle=270]{0225Feal.ps}
\includegraphics[height=\columnwidth,angle=270]{0225Oall.ps}

\caption{{\it{Top panel: }} The radial distribution of the Fe abundance. {\it{Bottom panel: }} The radial distribution of the O abundance. The Hydra~A data are from this paper, 2A~0335+096 from \citet{werner2006}, S\'ersic~159-03 from \citet{deplaa2006}, M87 from \citet{matsushita2003}, Centaurus from \citet{matsushita2007b}, Fornax from \citet{matsushita2007}, A~1060 from \citet{sato2007a}. The $r_{200}$ values were taken from \citet{reiprich2002} and scaled to $H_0=70 {\rm \:km\:s}^{-1} {\rm Mpc}^{-1}$. All the abundances are given with respect to the proto-Solar abundances by \citet{lodders2003}.}
\label{fig:abundprofall}
\end{figure}

\begin{figure}[!htp]
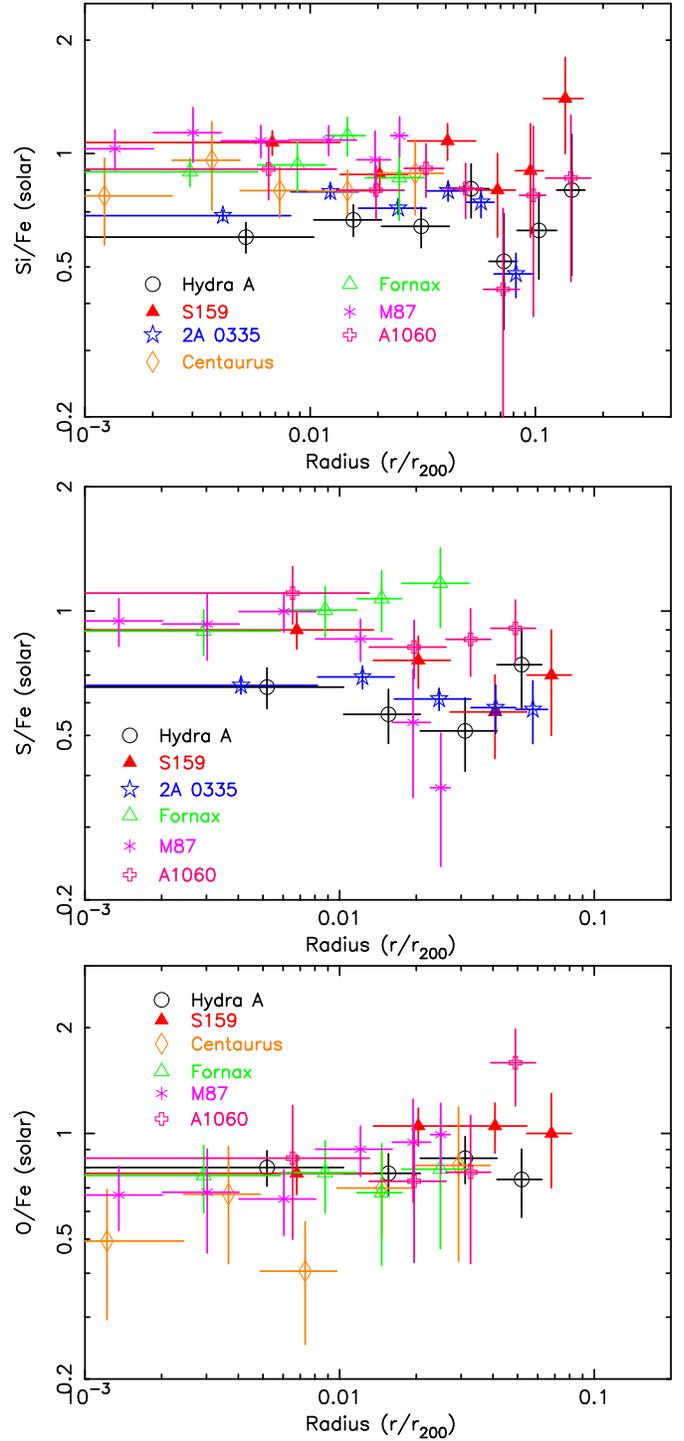

\includegraphics[height=\columnwidth,angle=270]{0225SiFe.ps}

\includegraphics[height=\columnwidth,angle=270]{0225SFe.ps}

\includegraphics[height=\columnwidth,angle=270]{0225OFe.ps}

\caption{{\it{Top panel: }} The radial distribution of the Si/Fe ratio. {\it{Middle panel: }}The radial distribution of the S/Fe ratio. {\it{Bottom panel: }} The radial distribution of the O/Fe ratio. }
\label{fig:relabunprof}
\end{figure}

In recent years, abundance profiles for several elements were determined for clusters which have deep observations with XMM-Newton and Suzaku. In Fig. \ref{fig:abundprofall}, we compare the radial distributions of the O and Fe abundances in Hydra~A with those found in other detailed spectroscopical analyses of clusters: 2A~0335+096 \citep[obtained by fitting the ``wdem'' differential emission measure distribution, no O abundances were presented;][]{werner2006}, S\'ersic~159-03 \citep[using the ``gdem'' model;][]{deplaa2006}, M87~\citep[using the 2T model within 2\arcmin\ and the MOS data;][]{matsushita2003}, Centaurus \citep[using the 2T model and MOS data;][]{matsushita2007b}, Fornax \citep[using the 2T model within 4\arcmin;][]{matsushita2007}, and A~1060 \citep{sato2007a}. The $r_{200}$ values were taken from \citet{reiprich2002} and scaled to $H_0=70 {\rm \:km\:s}^{-1} {\rm Mpc}^{-1}$. We excluded all abundance values determined with less than 3$\sigma$ significance.
The Fe abundance increases towards the center in all clusters, only Centaurus shows a drop of Fe in the innermost bin \citep[this is also the case for Perseus, not included in our sample, e.g.][]{sanders2004}. The Fe abundance in the Centaurus cluster is about a factor of 2 higher than in the other clusters, indicating a longer enrichment time scale.

In Fig. \ref{fig:relabunprof}, we compare the radial distributions of the relative metallicities of O, Si, and S with respect to Fe in all the above named data sets. The radial distribution of the relative abundance of Si/Fe is roughly constant within 0.1$r_{200}$ and the ratio is between $\sim$0.6 and $\sim$1.1 Solar. There seems to be a drop in Si/Fe in several data sets at $0.05r_{200}$, but the uncertainties of the Si/Fe ratio are generally large and they do not allow us to make definitive conclusions. The radial distribution of the S/Fe ratio shows a large scatter between the different clusters, and there is no obvious trend with radius. 

Interestingly, the observed trend with radius of the O/Fe ratio is much less pronounced than previously described by \citet{tamura2004}, who found a peaked distribution for Fe and a flat radial profile for O, leading to a marked increase in O/Fe with radius (up to O/Fe of 10 solar). Also, the trend of O with radius is clearly not flat, and there are indications of an increase in the O abundance towards the center for most clusters plotted in Fig. \ref{fig:abundprofall} (exceptions are the last data point in A1060 and a central drop in Centaurus, where Fe also shows a lower value in the center). In fact, taking into account all data points used in the plots and their respective error bars, we find that the O abundance significantly decreases with increasing radius, with a best-fit $d{\rm O}/d({\rm log}_{10}r/r_{200})=-0.48\pm0.07$, while the increase in the O/Fe ratio with radius is only less than 3$\sigma$ significant, $d({\rm O/Fe})/d({\rm log}_{10}r/r_{200})=0.25\pm0.09$. The Fe abundance does however decrease with radius significantly faster than the O abundance ($d{\rm Fe}/d({\rm log}_{10}r/r_{200})=-0.72\pm0.04$), suggesting that the slight trend in O/Fe is real. 

The interpretation of the flat O and peaked Fe profiles was, in previous publications, an early enrichment by \sncc\ in the protocluster phase, which led to the \sncc\ products being well mixed, and a later continuous enrichment by SN~Ia in the cD galaxy which continued to enrich the ICM on time scales much longer than \sncc, creating a peak in SN~Ia products at the cluster center. The flat Si/Fe ratio found in many clusters contradicted this picture. Si is produced by both SN~Ia and \sncc\ and according to this scenario its radial distribution should be shallower than that of Fe. This prompted \citet{finoguenov2002} to propose that there are two types of SN~Ia with different Si/Fe yields enriching the ICM. The authors proposed that, while the SN~Ia with higher Si yields and longer delay times contribute mostly within the elliptical galaxy, the SNe~Ia with lower delay times and lower Si yields would dominate the enrichment in the ICM at larger radii. 

However, Fig.~\ref{fig:abundprofall} shows that based on deep cluster observations with superior signal to noise ratio, there is evidence that the radial distribution of O is also centrally peaked. The O abundance determination with EPIC type CCDs is intrinsically very uncertain and sensitive to many systematics -- some of them already discussed above -- like background subtraction, calibration of the oxygen edge, $N_{\mathrm{H}}$, temperature determination, insufficient subtraction of the Galactic foreground emission which contains \ion{O}{viii} emission. This is especially important in cluster outskirts with low photon statistics. The results of \citet{tamura2004} are based on relatively short observations with XMM-Newton and their significances of the O abundance determinations in the outer parts of the individual clusters are typically only around $1\sigma$. We can conclude that the more accurate values shown in Figs. \ref{fig:abundprofall} and \ref{fig:relabunprof} suggest different O and O/Fe radial trends than previously assumed. 

We must therefore revise the picture that O/Fe significantly increases with radius and that the central abundance peak in clusters is created almost entirely by SNe~Ia and look for a mechanism that could create not only a peak in Fe and Si but also a peak in O. Sources of chemical enrichment in the ICM are SN~Ia, \sncc, and stellar winds. Slight increases in the relative abundances of O and Mg towards the center have been reported by \citet{bohringer2005}, who associate the presence of these peaks with stellar mass loss in the central galaxy. After a short incursion into what the observed abundance patterns can or cannot convey about supernova yield models, we will test this scenario for the case of Hydra A by evaluating the contributions from each source of chemical enrichment.

\subsection{ICM abundance patterns and supernova yield models}

\begin{figure*}
\includegraphics[height=\columnwidth,angle=270]{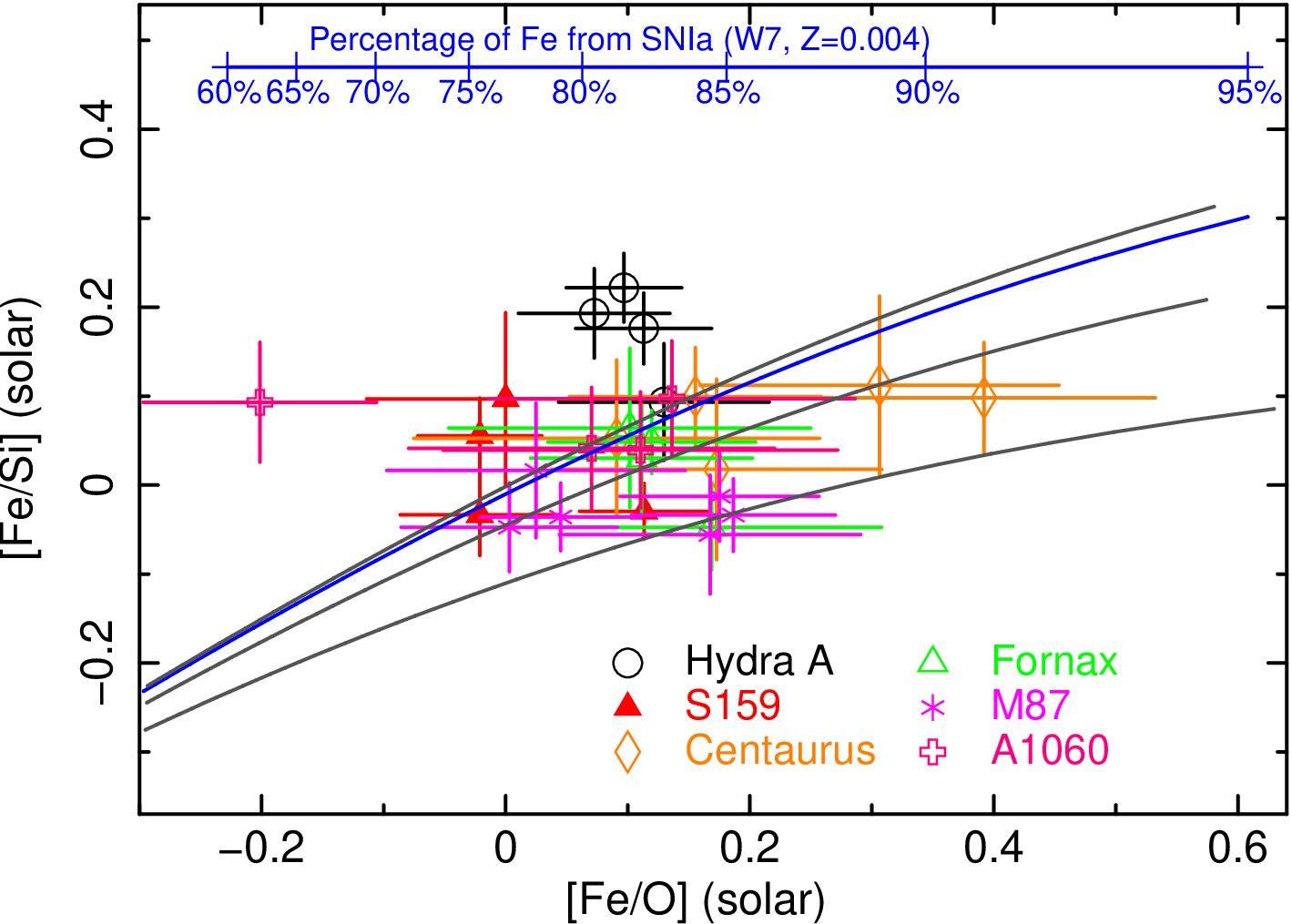}
\includegraphics[height=\columnwidth,angle=270]{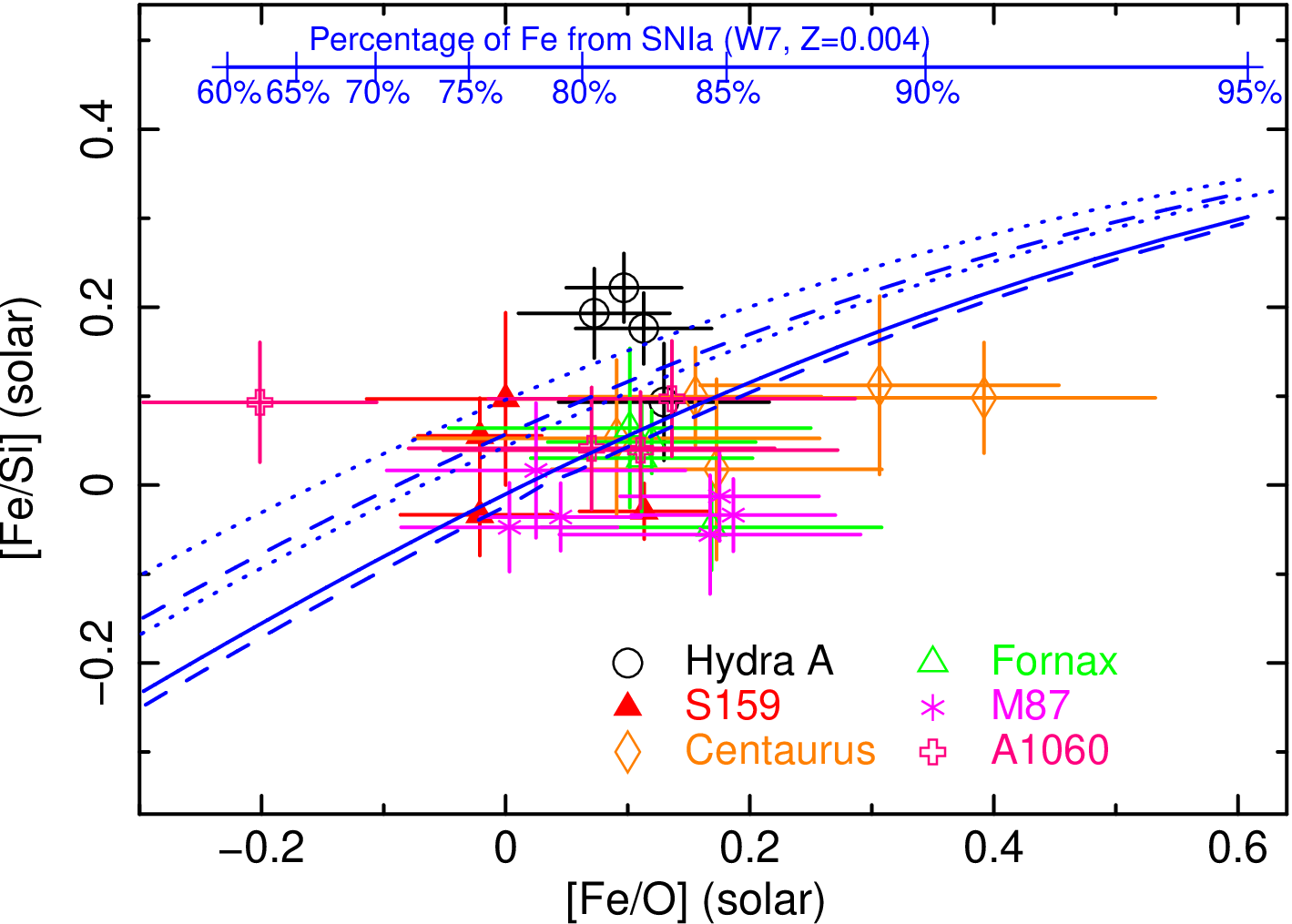}
\caption{The observed [Fe/Si] against [Fe/O] ratios (logarithmic values). {\it Left:} Expected trends for different SN~Ia models \citep[][WDD1-3 in gray increasing upwards and W7 in blue]{Iwamoto99}, assuming the \sncc\ yields from \citet{Kobayashi06} with Z=0.004 for the \sncc\ progenitor. {\it Right:} The effect on the expected trends when using different available \sncc\ models (SN~Ia yields were assumed to follow the W7 model). Dashed lines indicate the effect of using \sncc\ models with different Z's from \citet{Kobayashi06} (0. and 0.02, increasing upwards), dotted lines show trends using other published \sncc\ yields (above, \citealt{tsujimoto1995}; below, \citealt{nomoto2006}).}
\label{fig:sncontrib}
\end{figure*}

We plot in Fig. \ref{fig:sncontrib} the [Fe/Si]$\equiv$log$_{10}$(Fe/Si) against the [Fe/O] ratios for all data points available and over-plot the expected trends obtained by mixing an increasing amount of SN~Ia products into a gas whose initial composition is according to \sncc\ model yields. We use a variety of available SN~Ia models \citep[from][]{Iwamoto99} and \sncc\ models \citep{tsujimoto1995,nomoto2006,Kobayashi06}.

The amount of [Fe/Si] produced by different SN~Ia models decreases from WDD3 to W7, WDD2 and WDD1. This is shown by the four curves in the left panel of Fig. \ref{fig:sncontrib}, which all assume the same \sncc\ initial mass function (IMF) weighted average (between 0.07 and 50 $M_\odot$) yields from \citet{Kobayashi06} with an initial metallicity of the \sncc\ progenitor of Z=0.004 (0.2 solar). 
In the right panel of the same figure, we also show the influence on the expected [Fe/Si] vs. [Fe/O] curve if we assume different initial metallicities of the \sncc\ progenitor but the same SN~Ia model (in this case we chose to show this comparison for W7, which is most commonly used in galactic chemical evolution models). The two dashed line curves use the yields from \citet{Kobayashi06} with Z=0 (bottom) and Z=0.02 (top). Using the IMF weighted (10--50 $M_\odot$) \sncc\ yields of \citet{tsujimoto1995} and \citet{nomoto2006} \citep[for IMF-weighted values between 10 and 50  $M_\odot$, see][]{Werner08rev} would shift the model curves toward higher [Fe/Si].
 
We find a large diversity in the collected data, which do not strongly favor any particular SN~Ia model. Because the O/Fe radial trends are weak, as discussed above, there is no clear tendency for data points from larger radii to move along the plotted curves towards lower fractional contributions from SN~Ia. Rather, most data points are clustered at around 80\% of Fe coming from SN~Ia, which translates to a contribution of 30--40\% of SN~Ia by number.

Two interesting extreme examples are Hydra~A, whose very low Si abundance places it at the uppermost limit of [Fe/Si] which can be reached by any combination of SN~Ia and \sncc\ models considered, and M87 at the opposite end. The high Si abundance in M87 places it at the lower limit of Fe/Si that can be reproduced by currently available supernova models, and the data points here favor the WDD1 scenario. Most of the observed values for other clusters fall between the WDD1 and WDD2 model curves. An exciting puzzle is that, as Fig. \ref{fig:sncontrib} shows, a higher initial metallicity of the \sncc\ progenitor tends to shift the expected curve upwards to higher [Fe/Si]. Higher initial Z results in a higher [Ne/Fe] abundance ratio \citep{nomoto2006,Kobayashi06}. The RGS observations however show exactly the opposite trend: M87 which has the lowest [Fe/Si] has the largest [Ne/Fe] \citep[$1.40\pm0.11$ using a 2T model,][]{werner2006b} while Hydra~A with the highest [Fe/Si] has the lowest [Ne/Fe] ($0.73\pm0.18$ using a 2T model, this work).

Since supernovae are the most important sources of heavy elements, it should be possible to reconstruct the composition of any ICM by an appropriate fraction of SN~Ia to \sncc\ contributions. Understanding if the scatter of the data points in the [Fe/Si] vs. [Fe/O] plot has a physical meaning or is simply due to spectral fitting issues remains a challenge for future observational work, while trying to reproduce the diversity of abundance patterns we see can become a further input for supernova modeling. 

\subsection{The origin of the metal abundance peaks in Hydra A} \label{sect:peak}

In Sect. \ref{res:radprof}, we pointed out that the abundances of Fe, O, Si and S are larger in the core than in the cluster outskirts. We investigate here how these metal peaks can be produced. For this, we must first evaluate the mass of each metal needed to create the respective peak. This is given by:
\begin{eqnarray} \label{eq:mmet}
M_{\rm met}&=\sum_j&\gamma_{\rm met} \Delta Z_{{\rm met,}j} M_{{\rm gas,}j} \\
\end{eqnarray}
where $j$ is an index running over all annuli in the radial profile which are part of the peak, $\Delta Z_{{\rm met},j}= Z_{{\rm met},j}- Z_{{\rm met,out}}$ is the average metallicity excess between annulus $j$ and the value in the outskirts at which the profile flattens out, $M_{{\rm gas},j}$ is the gas mass inside annulus $j$, and $\gamma_{\rm met} $ is the mass fraction of the considered metal in the solar atmosphere, as determined from \citet{lodders2003}.

For each element, we assume as a ``base'' abundance $Z_{\rm out}$ (the abundance value outside the peak, to which the radial profiles flatten out) the value determined in the 3--8\arcmin\ region in Sect. \ref{global}. For O, the value in the outer region is not well constrained, therefore we will use the average of the two data points outside 2\arcmin\ where the O abundance could be determined in Sect.~\ref{res:radprof}. Our estimated ``base-abundances'' are thus $0.32$ solar for Fe, $0.15$ solar for Si, $0.10$ solar for S, and $0.20$ solar for O. If there is a continuing weak radial trend outside 3\arcmin\ for any of the elements \citep[see][]{Leccardi08}, these 3--8\arcmin\ values are an upper limit to the real abundances outside the peak. Given all the uncertainty sources involved, we show in this section only calculations based on the best-fit abundances without giving formal error bars. The presented results depend on the exact abundance profiles but also, among others, on the exact supernova yield and stellar wind modeling and on the IMF, star formation history and average stellar abundances in the central galaxy. We present estimates obtained using the best-guess assumptions based on available models for these parameters.

From Fig. \ref{fig:profile}, it can be seen that the abundance peak for all elements is clearly visible inside a radius of 2\arcmin\ (126 kpc), marked by a vertical dotted line in the plot. We thus sum Eq. \ref{eq:mmet} over the first three bins (0.--0.5\arcmin,0.5--1\arcmin, and 1--2\arcmin) in the radial profile to obtain $M_{\rm met}$ for each element.  $M_{\rm gas}$ can be determined from the spectrum normalization assuming a spherical or spherical shell geometry of the emitting region and using Eq. \ref{gasmass}. The corresponding metal masses in the peak are, in units of $10^8 M_\odot$, 2.7 for Fe, 1.9 for Si, 1.0 for S and 18.5 for O.

We assume that the enrichment of the metal abundance peak by the central galaxy took place over the last $10^{10}$ years (approximately between redshift $z=2$ and the present). This puts lower limits on the rates of metal production required to produce the peak, because mergers between $z=2$ and now could have dispersed some of the metals out towards the cluster outskirts. However, many clusters do require enrichment times of $\sim10^{10}$ years \citep{Boehringer04}, which may imply that mergers are not very efficient at disrupting cool cores.

According to \citet{Ciotti91}, the gas mass contributed by stellar winds as a function of time $t$ can be, in a simplified form which presupposes a Salpeter IMF \citep{Salpeter55} and single-age passively evolving stellar population, approximated as:
\begin{equation}\label{massloss}
\dot{M}_*(t) \approx 1.5\times10^{-11}L_B\:(t/t_H)^{-1.3}\:,
\end{equation}
where $L_B$ is the present-day blue band luminosity in units of $L_{B\odot}$ and $t_H$ is the Hubble time. The blue-band luminosity corresponding to the magnitude presented by \citet{Peterson86} for the central dominant galaxy is $9.2\times10^{10}L_{B\odot}$ within an aperture of 1.65\arcmin (108 kpc diameter at the redshift of Hydra~A, which likely includes the light emitted out to several times the galaxy's effective radius). Integrating this equation from $t(z=2)$ to the present ($t=t_H$), we obtain a gas mass contribution from stellar winds of $33.6\times10^9\:M_\odot$. Assuming the average metallicities of the stellar population to be solar, we estimate that the corresponding masses of Fe, Si, S and O ejected with the stellar winds are, respectively, 0.47, 0.28, 0.14, and 2.2 (in units of $10^8\: M_\odot$). We caution that the assumption of solar abundances for the stellar winds is a very uncertain one. Firstly, the stellar metallicities in the centers of central dominant galaxies can reach super-solar abundances. Secondly, elliptical galaxies have old stellar populations, meaning that their stars formed in the presence of less SN~Ia ejecta and the stellar composition should be closer to the \sncc\ abundance ratios. However, \citet{Kobayashi99} find an average [Mg/Fe] ratio in elliptical galaxies not very different from solar \citep[1.3 solar in the units of][]{lodders2003}. Moreover, cDs in cooling core clusters do show anomalous blue light emission and have star-forming regions \citep[see][for the case of Hydra~A]{McNamara95}. These young stars (which, being young, should have relatively strong stellar winds) must have incorporated much more SNIa products than the old populations of typical ellipticals.

If the rest of the Fe in the peak is produced entirely by SN~Ia, then assuming the yield from the WDD3 model of \citet{Iwamoto99}, which best reproduces the unusually high Fe/Si ratio in Hydra~A, $2.6\times10^8$ supernovae are required. For the W7 model, in turn, $3.0\times10^8$ supernovae would be needed.
On a time-scale of $10^{10}$ years, this implies an average rate of 2.6--3.0 SN~Ia per century. The current rate of SN~Ia in elliptical galaxies is  \citep{Cappellaro99}
\begin{equation}
\label{eq:ria}
R_{Ia}  = 0.18\pm0.06  \: ({\rm{100yr}})^{-1} \left(10^{10}L_{\odot} \right)^{-1},
\end{equation}
which implies 1.66 SN~Ia per century for the blue luminosity of the central galaxy in Hydra A, a factor of 1.6--1.8 too small compared to the required rate. There is however recent evidence that the SN~Ia rate in cluster ellipticals is larger than in field ellipticals. \citet{Mannucci08} find a SN~Ia rate in cluster ellipticals of $0.28^{+0.11}_{-0.08}  \: ({\rm{100yr}})^{-1} \left(10^{10}L_{\odot} \right)^{-1}$, in excellent agreement with the rate required in Hydra A. Furthermore, the rate of SN~Ia may have been larger in the past. \citet{Renzini93} propose a time-dependence of $(t/t_H)^{-k}$ with $k$ spanning 1.1 up to 2. Integrating this between $t(z=2)$ and now, the average SN~Ia rate over the past $10^{10}$ years could be 2 to 4.1 times greater than the present one, and could have produced the observed amount of Fe in the peak over the assumed enrichment time.

Using the yields from the WDD3 model, $2.6\times10^8$ supernovae would produce $0.41\times10^8 \: M_\odot$ of Si, 0.25 of S and 0.15 of O. Adding the contributions from SN~Ia and stellar winds therefore, we still lack $16.2\times10^8 \:M_\odot$ of O, $1.2\times 10^8\:M_\odot$ of Si, and $0.6\times 10^8\:M_\odot$ of S compared to the estimated total masses of these elements in the central peak. Even with the W7 SN~Ia model, which produces the highest amount of O among the models of \citet{Iwamoto99}, still $15.9\times10^8 \:M_\odot$ of O would be unaccounted for solely by adding the contributions of SN~Ia and stellar winds. 

One possible solution is to consider that the metal input by stellar winds is larger than we previously assumed, either because of super-solar metallicities in the central galaxy or a higher stellar mass loss rate than predicted by models (due, for example, to a more complex star formation history than assumed). Keeping the assumption of solar ratios, if the stellar winds would produce around 8 times more of each metal than assumed above, meaning that a correspondingly lower number of SN~Ia would be needed to create the Fe peak, we could reproduce well the observed masses of O, Si and S in the peak both assuming W7 and WDD3 SN~Ia yields. Without the assumption of solar ratios but keeping the solar Fe abundance in the stellar winds and the mass loss rate calculated in Eqn. \ref{massloss} does not present a viable alternative, since the Si/Fe and S/Fe in the stellar winds would need to be 5.3--5.4 solar, more than is reproduced by any \sncc\ model. As a combination of the two possibilities, if the stellar winds would produce only 3 times more Fe than assumed above but would have Si/Fe and S/Fe of 2 solar and O/Fe of 2.8 solar, we could also reproduce the observations well. 

An alternative for creating the additional O, Si and S is to consider a contribution from \sncc\ to the metal peak. Assuming an unchanged stellar mass loss rate and average stellar wind metallicity and using the initial mass function weighted average \sncc\ yields from \citet{Kobayashi06}, approximately $8\times10^8$ core-collapse supernovae would be needed, in addition to the contribution of SN~Ia and stellar winds, to reproduce the observed O peak. For Si and S, only 6 and 3$\times10^8$ \sncc\ are needed, respectively. This would mean either that there was a significant contribution by \sncc\ in the central galaxy over the last $10^{10}$ years (note that almost twice as many \sncc\ than SN~Ia are required!), or if the initial enrichment by \sncc\ in the protocluster phase was not as well mixed on large scales as previously thought, and some peak in the distribution of \sncc\ products existed prior to further enrichment with predominantly SN~Ia elements. Both cases provide alternatives to the approach of \citet{finoguenov2002} for explaining why the radial distribution of Si is not shallower than that of Fe, as would be expected if only SN~Ia contributed to the central abundance peak.

Additionally, according to recent simulations \citep{Domainko06}, ram-pressure stripping of cluster galaxies also leads to a stronger enrichment of the cluster centers compared to the outskirts, and could lead to comparable Fe, Si, S and O peaks. The typical spatial scale of this enrichment however ($\sim$ 1 Mpc) is considerably larger than the extent of the peak in Hydra A ($\sim$ 130 kpc).

\subsection{Gas uplift by the AGN and metal transport into the ICM}
\label{metaltransp}

As we point out already in Section \ref{sect:2Dmaps}, there is a remarkable correlation between the 327 MHz large-scale radio lobes, cool arm-like extensions in the temperature map, and metal-rich filaments in the Fe abundance map. This suggests that buoyant radio bubbles produced by the AGN uplift cool, metal-rich gas from the central parts of the X-ray halo, thereby contributing to the transport and distribution of heavy elements produced in the central galaxy into the ICM \citep{Churazov01,simionescu2007b}. 

The metallicity map shows a clear elongation along the N--S direction, with filaments coinciding very well with low-temperature features in the central parts of the temperature map (Fig. \ref{fig:tmap}). Similar elongations in the metallicity map are seen in simulations of AGN-induced metal-transport in galaxy clusters \citep{Roediger06}.
Towards the North, the cool, metal-rich gas overlaps well with the rising stem of the northern radio-lobe. Towards the South, the low-temperature feature is oriented at a slightly more easterly angle than the southern radio lobe, and is associated with a weaker increase in metallicity than in the N. The southern radio lobe is bent and clearly more disturbed than the northern lobe, and the same process which leads to the deformation of the radio plasma could also account for the displacement between the radio emission and the cool filament.

\citet{simionescu2007b} used the fact that the gas uplifted by the AGN in M87 is multi-phase to infer the metallicity, mass, origin and chemical enrichment history of this gas. Unfortunately, in Hydra A, the EPIC spectra of the cool arm-like features in the temperature map do not require additional temperature components. This can be explained if the amount of cool gas which has a temperature around and below $\approx$1 keV, where the shape of the Fe-L complex becomes clearly different and can indicate multi-phase structure, is very low. Based on the Fe-L shape, one of the very few spectral indications for multi-temperature accessible with EPIC, gas above $\approx$2 keV is difficult to distinguish from gas in hotter phases. Indeed, as shown in Section \ref{section:RGS} where we do detect a small amount of cool gas below 1 keV, the spectrum normalization of this gas is only 1.2--1.4\% of the total emission of the hot halo, which is impossible to uniquely identify using the EPIC cameras. 

Based on the RGS result, we can try to infer some further properties of the cool gas. The RGS extraction region we used was 3\arcmin\ wide in the cross dispersion direction and effectively $\sim$10\arcmin\ long in the dispersion direction. If we assume that the cool gas is concentrated in the cluster center, thus within a 3\arcmin-diameter circular region which is the 2D-projection of a sphere with a radius of 1.5\arcmin\ (94.5 kpc), then the mass of the cool gas corresponding to its emission measure of $2.6\times10^{65} {\rm \: cm}^{-3}$ would be $1.8\times10^{11} M_\odot$. The mass is determined using:
\begin{eqnarray}\label{gasmass}
Y = \int n_{\mathrm{e}} n_{\mathrm{H}} dV \approx n_{\mathrm{e}} n_{\mathrm{H}} V \approx 1.2\: n_{\rm H}^2 V \nonumber\\
M = \sum_i m_i n_i V\approx (m_{\rm H} n_{\rm H} + m_{\rm He} n_{\rm He})V\approx 1.4 m_{\rm H} n_{\rm H}V
\end{eqnarray} 
where $n_{\mathrm{e}}$, $n_{\mathrm{H}}$ and $n_{\rm He}$ are the electron, proton and helium number densities, respectively, $m_{\rm H}$ is the proton mass, $m_{\rm He} = 4 m_{\rm H}$, and $V$ is the volume of the emitting region. Approximately, $n_{\rm e} \approx 1.2\: n_{\rm H}$, and assuming a solar He/H fraction, $n_{\rm He} / n_{\rm H} \approx 9.5\%$. We neglect the mass contribution of ions heavier than He which are scarce compared to H and He. 

\begin{figure}
\includegraphics[width=\columnwidth]{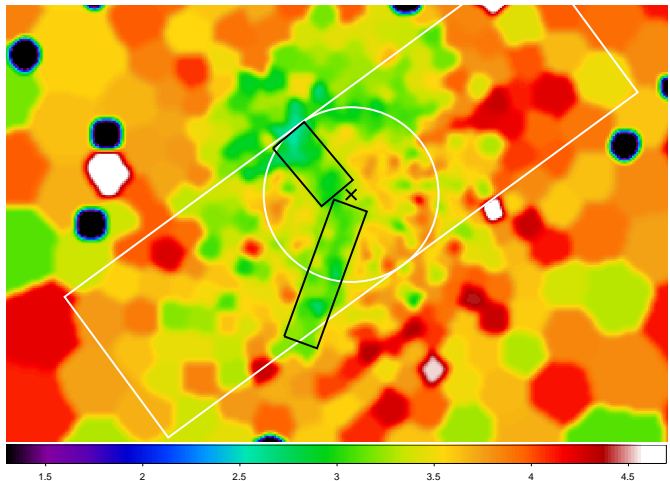}
\caption{Zoom-in on the temperature map in Fig \ref{fig:tmap} smoothed with a $16^{\prime\prime}$ Gaussian filter. A circle with radius of 1.5\arcmin\ and the 3 by 10\arcmin\ RGS extraction region are overlaid in white. The cluster center is marked with a black cross and the black rectangles mark, in projection, the cylinders by which we approximate the volume of the cool ``arms''. Colorbar units are keV.}
\label{fig:tmapreg}
\end{figure}

If we assume on the other hand that the cool gas comes only from the cool ``arms'' seen in the maps and not from the entire 1.5\arcmin-radius sphere, we can approximate the volume of this gas by a sum of cylinders for the Northern and Southern ``arms'', depicted in Fig. \ref{fig:tmapreg}. Both cylinders are bound to lay inside the 3 by 10\arcmin\ RGS extraction region. The N cylinder has a radius of 0.35\arcmin\ and a length of 1.3\arcmin, the S cylinder has a radius of 0.3\arcmin\ and is 2.5\arcmin\ long. This yields a volume of $8.9\times10^{69} \:{\rm cm}^3$, and a mass of $5.1 \times 10^{10} \:M_\odot$. This is only a rough estimate because it assumes the cold gas to be distributed uniformly in the considered cylindrical regions.

We furthermore need to take into consideration the possibility that the cool gas may not fill the entire assumed volume of the ``arms'', but may be concentrated in thin filaments which occupy only a fraction (so-called filling factor) of this volume. Since these filaments are unresolved, the only way to estimate this filling factor is to assume pressure equilibrium between the cool and hot phases, in the form $n_{\rm c} kT_{\rm c} = n_{\rm h} kT_{\rm h}$ which, using $n_{\rm c,h} \propto \sqrt{Y_{\rm c,h}/V_{\rm c,h}} $, implies
\begin{equation}
\frac{V_{\rm c}}{V_{\rm h}} = \frac{Y_{\rm c}}{Y_{\rm h}}\left(\frac{T_{\rm c}}{T_{\rm h}}\right)^2
\label{eq:ff}
\end{equation}
where c and h are subscripts denoting the properties of the cool and hot gas determined by RGS, respectively. From Eqn. \ref{eq:ff}, we obtain $V_{\rm c}/V_{\rm h}=1.0\times10^{-3}\approx V_{\rm c}/V_{\rm tot}$, which leads to a mass of cool gas of $1.6\times10^9 \: M_\odot $. This is a factor of $\sim$3 larger than the mass of uplifted cool gas calculated in M87 by \citet{simionescu2007b}.

If we assume that the average displacement of the mass of cool gas determined above is on the order of 1\arcmin, which is about 63 kpc, and use the integrated mass profile presented by \citet{David01}, we can estimate the gravitational energy needed to uplift this gas at $3.3\times10^{58}$ ergs, which is 8 times greater than the energy needed for gas uplift in M87 but nevertheless very small compared to other processes related to AGN-ICM interaction in Hydra A, which require approximately $10^{61}$ ergs (large-scale shock, \citealt{Nulsen05,Simionescu08shock}, and cavities, \citealt{Wise07}). 

Finally, we estimate the amount of Fe which is being transported by the AGN through the uplift of the cool gas. Assuming a gas mass for the cool gas of $1.6\times10^9 \:M_\odot $ and a metallicity, as determined from RGS, of 0.35 solar, and using $M_{\rm Fe}=\gamma_{\rm Fe} M_{\rm gas}$, this gives $M_{\rm Fe}=7.9\times 10^5 M_\odot$. However, this is almost certainly too low, since the uplifted cool gas should be more metal-rich than the ambient gas, as the structure in the Fe map suggests, while the abundances of the hot and cool gas were coupled in the RGS fit. Moreover, some of the uplifted metal-rich gas may have higher temperatures, therefore its mass is not included in our estimate of $1.6\times10^9 \:M_\odot $ of uplifted gas, which is based solely on the RGS best-fit normalization of the 0.62 keV gas. This leads to an underestimation of the mass of uplifted gas and implicitly of the mass of uplifted Fe. 

Another way to estimate the mass of transported Fe is to assume a typical metallicity outside the metal-rich regions and apply Eq. \ref{eq:mmet} summing over all bins which lie inside the rectangles by which we approximated the projected shape of the N and S arms in Fig. \ref{fig:tmapreg}. Assuming the average metallicity outside the metal-rich regions to be 0.445 solar (the average in the inner 3\arcmin, Sect. \ref{global}), the mass of uplifted Fe would be $1.7\times10^7\:M_\odot$, while if the metallicity outside the metal-rich regions is 0.32 solar (the ``base'' abundance for Fe discussed in Sect. \ref{sect:peak}), we obtain an upper limit of $6.9\times10^7\:M_\odot$ for the mass of Fe transported by the AGN together with the uplifted gas. 

In turn, the total mass of Fe in the central 0.5\arcmin\ region is $11.5\times10^7\:M_\odot$, meaning that the AGN transported about 15\% and up to 60\% of the amount of Fe currently present in the inner part of the Hydra A halo out to larger radii by uplifting central, metal-rich gas. We note that the O/Fe value determined by extracting a spectrum in the N arm ($0.80\pm0.14$) is consistent with the average O/Fe ratio in the inner 3\arcmin\ (Table \ref{tab:global}). This means, in agreement with \citet{simionescu2007b}, that the AGN, at least at this advanced stage of the evolution of the galaxy, transports metals into the ICM without altering the relative abundance patterns.

We already noted that some of the uplifted gas may have temperatures higher than 0.62 keV, therefore its mass is not included in our estimate of $1.6\times10^9 \:M_\odot $, which results in underestimating the uplifted gas mass. We showed a method of estimating the uplifted mass of Fe independent of the estimated mass of the 0.62 keV component. If we then assume an average metallicity for the uplifted gas of 2 solar \citep[as in the case of M87,][]{simionescu2007b}, we can compute the total mass of uplifted gas to be $6.1\times10^9 \:M_\odot $, corresponding to an uplift energy of $1.25\times10^{59}$ ergs. The results for different methods of estimating the gas and Fe mass and the uplift energy are summarized in Table \ref{tab:mass_gas_fe}.

\begin{table}
\caption{Uplifted gas mass, Fe mass, and energy, $^{*)}$ with equal metallicities of the 0.62 keV and the hot gas from the RGS fit, $^{**)}$ if the average metallicity of the total uplifted gas is 2 solar.}
\begin{center}
\begin{tabular}{l|cccc}
\hline\hline
 & gas mass ($M_\odot $) & Fe mass ($M_\odot $) & energy (ergs)\\
 \hline
0.62 keV	 & $1.6\times10^9$      &$7.9\times10^{5\:*)}$    &  $3.3\times10^{58}$     \\
total 	           & $6.1\times10^{9\:**)}$ &$1.7\times10^7$ & $1.25\times10^{59}$	\\

\hline
\end{tabular}
\label{tab:mass_gas_fe}
\end{center}
\end{table} 

\section{Conclusions}

We analyzed a deep $\sim120$ ks XMM-Newton exposure of the cooling core cluster of galaxies Hydra~A. We extracted spectra from two large regions in the cluster core and in the outskirts to study the global properties of the cluster with the best statistics possible. We also analyzed the RGS spectrum extracted from a 3 by 10\arcmin\ region. We find that
\begin{itemize}
\item
the shape of the Fe-L complex in the EPIC spectrum does not show strong indications for the presence of cooler gas as observed in other cooling core clusters. However, a multi-temperature model is needed to simultaneously fit both Fe-L and Fe-K lines appropriately. The best available model achieving this is a Gaussian distribution of the emission measure ({\it gdem}) around the best-fit average temperature.
\item
the best-fit ({\it gdem}) model shows a very broad distribution, with a full-width at half-maximum of 4.2~keV. We also fitted polynomial emission measure distributions which confirm the broad shape of $dY/dT$.
The distribution is significantly broader than expected from the substructures visible in the 2D map in the considered region, suggesting the presence of intrinsic multi-temperature structure in each bin of the temperature map. We suggest that this multiphase structure can be due in part to the projection of shocked gas in front of and behind the cluster center, in part to the projection of cooler gas from the cluster outskirts along the line of sight, and in part to dense, unresolved cool gas blobs in the considered extraction region.
\item
we can accurately determine abundances for 7 elements in the cluster core with EPIC (O,Si,S,Ar,Ca,Fe,Ni). In the cluster outskirts, only Si, S and Fe abundances are determined with better than 3$\sigma$ significance and the large errors do not enable us to draw conclusions about possible differences in Si/Fe and S/Fe with respect to the cluster center.
\item
the abundances of 3 elements (O,Ne,Fe), one of which is inaccessible with EPIC, are determined with RGS. The O/Fe ratios from EPIC and RGS are consistent. While no multi-temperature structure beyond the {\it gdem} model can be constrained with EPIC, the RGS fit requires a cool component with a temperature of $0.62\pm0.04$ keV and a normalization of only 1.2\% of the hot ambient. This gas is detected with a significance of 6.5$\sigma$. 
\item
the Gaussian emission measure distribution model gives lower Fe abundances than a single temperature model, which can explain why simulations show that the best-fit Fe abundance in clusters with intermediate temperatures is over-estimated.
\end{itemize}
We also determined temperature and abundance profiles from seven annuli and compared our results with radial profiles of other clusters for which deep observations and detailed chemical enrichment studies are available. We show that
\begin{itemize}
\item
the abundance profiles for Fe, Si, S, but also O are centrally peaked in Hydra A.
\item
the radial profiles of the O abundance are peaked also in other clusters for which deep data are available. The increase in O/Fe with radius is very small. Combining the Hydra A results with 5 other clusters for which a detailed chemical abundance study has been performed, we find $d{\rm O}/d({\rm log}_{10}r/r_{200})=-0.48\pm0.07$, while the increase in the O/Fe ratio with radius is only less than 3$\sigma$ significant, $d({\rm O/Fe})/d({\rm log}_{10}r/r_{200})=0.25\pm0.09$. 
\item
stellar winds and the chemical enrichment by the number of type Ia supernovae needed to produce the Fe peak in Hydra A do not reproduce the estimated O, Si and S peaks. For this, either the amount of metals produced by stellar winds would have to be 3--8 times higher than predicted by available models or $3-8\times10^8$ \sncc\ would be needed in addition to the contribution from stellar mass loss and SN~Ia. Possibly, the initial enrichment by \sncc\ in the protocluster phase was not as well mixed on large scales as previously thought, and some peak in the distribution of \sncc\ products existed prior to further enrichment with predominantly SN~Ia elements.
\item
mainly because of the low Si abundance, Hydra A requires either a WDD3 or W7 SN~Ia model to reproduce the observed relative abundance patterns. Most other clusters lie between the WDD1 and WDD2 models. A 30--40\% contribution by SN~Ia compared to \sncc\ (by number) is needed to reproduce the observed relative abundance patterns at all radii (below 0.1 $r_{200}$) in all the considered clusters. 
\item
the best-fit Galactic absorption column density, $N_{\mathrm{H}}$, when fitting the full energy band, is lower than the value determined from \ion{H}{i} data and has a minimum on the cluster center, which may be interpreted as inverse Compton or soft excess emission. The best-fit $N_{\mathrm{H}}$ in the 0.35--2 keV band however is in agreement with the \ion{H}{i} data, dismissing this possibility.
\end{itemize}
Finally, we produced 2D temperature and metallicity maps of Hydra A. From these we can conclude that
\begin{itemize}
\item
the temperature map shows cooler gas extending in arm-like structures towards the north and south. The cool gas structures, and especially the northern one, appear to be richer in metals than the ambient medium and spatially correlated with the large-scale radio lobes. The northern ``arm'' is associated with a bright 1\arcmin\ long filament seen in the Chandra image.
\item
based on the geometry of the cool ``arms'' seen in the temperature map and on the best-fit normalization of the 0.62 keV component in RGS, the estimated mass of cool gas, which was probably uplifted by the AGN to create these structures, is $1.6\times10^9\:M_\odot$. The energy needed for this uplift is $3.3\times10^{58}$ ergs, which is 8 times greater than the energy needed for gas uplift in M87 but nevertheless very small compared to other processes related to AGN-ICM interaction in Hydra~A (large-scale shock and cavities), which require approximately $10^{61}$ ergs. 
\item
the best estimate of the mass of Fe uplifted together with the cool gas is $1.7\times10^7\:M_\odot$, 15\% of the total mass of Fe in the central 0.5\arcmin\ region. If the average metallicity of the uplifted gas is 2 solar, as in M87, the total mass of uplifted gas (not only the 0.62 keV component) is $6.1\times10^9\:M_\odot$ and the uplift energy is $1.25\times10^{59}$ ergs.
\item
the O/Fe value in the N arm is consistent with the average O/Fe ratio in the inner 3\arcmin. The transport of metals by the AGN thus presently does not alter the relative abundance patterns.
\end{itemize}

\begin{acknowledgements}
We would like to thank W. R. Forman and J. Vink for helpful discussions. We acknowledge the support by the DFG grant BR 2026/3 within the Priority Programme "Witnesses of Cosmic History" and NASA grant NNX07AQ18G 16610022. AS would like to thank the Harvard-Smithsonian CfA, SRON, and Jacobs University Bremen for their hospitality. AF acknowledges support from BMBF/DLR under grant 50 OR 0207 and MPG.
This work is based on observations obtained with XMM-Newton, an ESA science mission with instruments and contributions directly funded by ESA member states and the USA (NASA). The Netherlands Institute for Space Research (SRON) is supported financially by NWO, the Netherlands Organization for Scientific Research. 
\end{acknowledgements}

\bibliographystyle{aa}
\bibliography{clusters,bibliography}

\end{document}